\documentclass[pdflatex,sn-mathphys-num]{sn-jnl}

\usepackage{CJKutf8}
\usepackage{amsmath,amssymb,amsthm}
\let\orcidlogo\relax
\usepackage{orcidlink}

\DeclareMathOperator{\Ric}{Ric}
\DeclareMathOperator{\tr}{tr}
\DeclareMathOperator{\spec}{spec}
\newcommand{\dd}{\mathrm{d}}

\hypersetup{hidelinks}
\numberwithin{equation}{section}

\theoremstyle{thmstyleone}
\newtheorem{theorem}{Theorem}[section]
\newtheorem{proposition}[theorem]{Proposition}
\newtheorem{lemma}[theorem]{Lemma}
\newtheorem{corollary}[theorem]{Corollary}

\theoremstyle{thmstylethree}

\theoremstyle{remark}
\newtheorem{remark}[theorem]{Remark}

\begin{document}

\title{Smooth Rigidity Near Extremal Horizons}

\author*[1]{\fnm{Ye} \sur{Zhou}~(\begin{CJK*}{UTF8}{gbsn}周烨\end{CJK*})\,\orcidlink{0009-0004-7050-7736}}
\email{ye.zhou.horizon@gmail.com}

\affil*[1]{\orgname{Independent Researcher},
  \orgaddress{\city{Kunshan}, \state{Jiangsu}, \country{China}}}

\abstract{
The Einstein equations constrain the transverse Taylor series at a degenerate Killing horizon, but even complete knowledge of that series need not determine a smooth spacetime: a nonzero difference can vanish to every order at the horizon. We show that this ambiguity disappears when the limiting metric on the space of Killing orbits admits a spacelike graph over a compact horizon section. In dimensions $d\geq4$, smooth vacuum Einstein germs with the same marked transverse jets are then locally isometric on a two-sided neighborhood of the whole section. The proof combines a terminal wave-map gauge with energy estimates on the orbit cylinder and yields a finite weighted matching criterion. For unwarped horizons with positive-Einstein sections, a condition on the full Lichnerowicz spectrum reduces the smooth classification to parallel first transverse data. All such data have convergent multiwarped Einstein realizations, including examples whose sections have negative sectional-curvature directions. We also obtain a finite-regularity version for round sections. The continuation argument extends to four-dimensional Einstein--Maxwell theory and gives smooth charged-Nariai rigidity. Finally, rotating Nariai data admit the required spacelike graph throughout the regular branch; an all-order tensor estimate yields smooth Kerr--de Sitter or rotating-Nariai rigidity for $a^2/r_0^2\leq1/8$, without assuming ambient axial symmetry.
}


\maketitle

\section{Introduction}\label{Sec:Intro}

An extremal horizon gives a hierarchy of geometric data. The intrinsic metric and rotation form satisfy the near-horizon equations, while transverse differentiation of the Einstein equations provides further constraints. Even if those constraints determine every transverse derivative, however, a nonzero smooth difference may vanish to infinite order at the horizon. Classification of near-horizon geometries, uniqueness of formal expansions and uniqueness of smooth spacetime germs are therefore different questions. Intrinsic rigidity results have considerably sharpened our understanding of the first of these \cite{KunduriLucietti2013,DunajskiLucietti,CollingKatonaLucietti,Colling2025,Lucietti2026}. We address the last.

The distinction appears clearly in the work of Katona and Lucietti \cite{KL}, who classified analytic vacuum spacetimes with compact maximally symmetric static extremal horizons as locally Schwarzschild--de Sitter or Nariai. Katona proved an Einstein--Maxwell analogue \cite{Katona2024}. In both cases the transverse equations determine the Taylor series, after which analyticity supplies the passage to a genuine metric. First-order deformation studies \cite{LiLucietti2016,LiLucietti2019,Kolanowski2022} and the general theory of null-hypersurface expansions \cite{MarsI,MarsII} clarify the transverse equations, but leave the possibility of a smooth flat remainder. Removing that remainder is the starting point of this paper.

Our continuation theorem has a simple geometric hypothesis. Let $S$ be a compact spacelike section of a degenerate Killing horizon, with a fixed marking and nonvanishing normalized generator, and denote its Gaussian-null data by $(F_0,h_0,\gamma)$. We require one globally defined smooth function $\psi$ on $S$ such that $F_0>|h_0+d\psi|_\gamma^2$. Geometrically, this means that the limiting Lorentzian metric on the space of Killing orbits admits a compact spacelike graph. The graph need not be a constant-radius slice. Under this condition, Theorem~\ref{strict:theorem} shows that two smooth vacuum Einstein germs with identical marked transverse jets agree, up to a marked isometry, on a two-sided neighborhood of the entire section. The proof turns transverse flatness into decay faster than any exponential on the orbit cylinder, then excludes such decay through a terminal wave-map gauge and energy estimates. It also gives a finite weighted matching theorem for fixed limiting data and coefficient bounds (Theorem~\ref{thm:finite-graph}). For round sections, a separate finite-decay argument extends the classification to sufficiently high finite metric regularity. Unlike continuation arguments at bifurcate or nondegenerate horizons \cite{CMStaticity,PetersenWave,Petersen,KP}, this construction uses the degenerate horizon itself.

To obtain classifications, we combine continuation with the transverse equations. For an unwarped horizon with a compact positive-Einstein section, the relevant operator is the Lichnerowicz Laplacian on the full trace-free tensor space. If its first kernel consists precisely of parallel tensors and its higher indicial shifts are nonresonant, the first transverse coefficient determines all subsequent ones. Every permitted parallel coefficient has an explicit convergent multiwarped realization, and these exhaust the smooth germs (Theorems~\ref{nr:realization} and \ref{nr:smooth}). The criterion is spectral, not a sectional-curvature sign condition. In particular, the K\"ahler--Einstein identity of Dai, Wang and Wei \cite{DWW2007}, together with known geometry and existence results \cite{KotschickThung2020,LiuXu2019,Mok1988}, gives rigid sections with negative sectional-curvature directions, including $G_2/U(2)_+$ and smooth cubic threefolds.

For rotating Nariai data, the two requirements separate more sharply: a spacelike orbit graph exists throughout the regular cosmological Kerr--de Sitter double-root branch, whereas the full transverse operator is uniformly coercive for $a^2/r_0^2\leq1/8$. In that interval, every smooth vacuum germ with these horizon data is locally Kerr--de Sitter or rotating Nariai, without assuming axial symmetry of the ambient metric (Theorem~\ref{rotating:ambient-classification}). This goes beyond the first-order and partial infinitesimal analyses in \cite{JezierskiKaminski2013,BahuaudEtAl2024}; the smaller interval comes from the coercivity estimate, not the orbit geometry.

The method also extends to four-dimensional Einstein--Maxwell fields invariant under the horizon generator. Retaining the electromagnetic source in the twist equation yields smooth coupled jet determination and, using the known transverse recursion, charged-Nariai rigidity for $1/2<\Lambda R^2\leq1$ (Theorem~\ref{em:jet} and Corollary~\ref{em:charged-nariai}). All these statements concern a neighborhood of a marked horizon section; they require neither the asymptotic conditions of global uniqueness problems \cite{ChruscielCostaHeusler2012,ChruscielNguyen2010} nor the bifurcate subextremal setting of recent Kerr--de Sitter rigidity \cite{Fang2025}.

Section~\ref{sec:horizon-orbit} develops the orbit geometry, and Section~\ref{sec:unique-continuation} proves the continuation theorem. Section~\ref{sec:transverse-rigidity} treats the spectral classification and its realizations; Sections~\ref{sec:einstein-maxwell} and \ref{sec:rotating-nariai} give the coupled and rotating applications. Section~\ref{Sec:Discussion} discusses the remaining questions.

\section{The Horizon and Its Orbit Space}\label{sec:horizon-orbit}\label{sec:geometry}

We normalize the vacuum equation as $\operatorname{Ric}(g)=\Lambda g$ and write $m=d-2$. Let $S$ be a compact connected spacelike section of a degenerate Killing horizon, with a nowhere-vanishing generator $K$ and a fixed identification with a reference manifold. The distinction between data intrinsic to a horizon and data describing its transverse neighborhood is fundamental here; see \cite{AshtekarBeetleLewandowski2002,LewandowskiPawlowski2003,KunduriLucietti2013,Jezierski2009}. The other null normal $N$ is determined by $g(N,K)=1$, $g(N,N)=0$, and $g(N,TS)=0$. The Killing flow and affine null geodesics with initial tangent $N$ give stationary Gaussian-null coordinates
\begin{equation}
g=2\,dv\,dr+r^2F\,dv^2+2rh_a\,dv\,dx^a+ \mu_{ab}\,dx^a\,dx^b,\qquad K=\partial_v. \label{pos:gnc}
\end{equation}
The coefficients are independent of $v$. The same $(v,r)$ is used on all section charts. Compactness provides a common collar width; only a local interval of Killing flow is needed. When translating the cut by a bounded function on $S$, one can first extend these stationary coefficients to the auxiliary product $\mathbb R_v\times(-\epsilon,\epsilon)_r\times S$. This local extension is Einstein and introduces no completeness assumption on the original spacetime.

For the unwarped vacuum results the horizon data are
\begin{equation}
F(0)=\Lambda>0,\qquad h(0)=0,\qquad \mu(0)=\gamma, \qquad \operatorname{Ric}(\gamma)=\Lambda\gamma. \label{eq:unwarped}
\end{equation}
The last equation follows from the first three and the horizon Einstein equations \cite{KunduriLucietti2009,DKLS,BahuaudEtAl2022}. No sectional-curvature hypothesis is implicit in these data. Matching transverse jets means equality of $\partial_r^j(F,h,\mu)|_{r=0}$ as tensors on the identified section, in the Gaussian-null constructions just specified. Equivalently the two tensor germs agree to infinite order in this fixed transverse identification. It is preserved by applying the same change of marking and generator normalization to both germs. An independent supertranslation of one cut changes the data and is not silently quotiented out in a marked comparison.

The actual-section hypothesis can be recorded independently of Einstein's equation.  Let $S$ be an embedded compact spacelike codimension-two submanifold of a smooth Killing horizon, transverse to its nonzero generator $K$.  The two equations $g(N,K)=1$, $g(N,TS)=0$, together with $g(N,N)=0$, determine one smooth transverse null vector $N$ on $S$. The flow-box theorem applied to $K$ gives a local horizon map $(-\delta,\delta)\times S\to\mathcal H$.  Compactness supplies a uniform $\delta$ and prevents collisions after shrinking it: otherwise a sequence of arbitrarily short collisions would converge to a violation of a local flow box at some point of $S$.  Applying the geodesic exponential to the transported $N$ similarly gives a diffeomorphism from a uniform $(-\delta,\delta)_v\times(-\epsilon,\epsilon)_r\times S$ onto a collar. It is not necessary to choose angular coordinates globally; the tensor coefficients agree under the original section transition maps.  Killing invariance, affine normalization and degeneracy give exactly the Gaussian-null form in the text.

An isometry fixing $S$, its marking and $K$ also fixes $N$.  If its first differential is the identity along the horizon, it sends each transverse affine null geodesic to the one with the same initial data. Geodesic uniqueness then identifies $r$ and all the section labels, so in the two common Gaussian-null constructions it is the identity.  This is why the final conclusion is stronger than an unspecified local isometry.

A supertranslation to the cut $v=f(x)$ changes this normalization and is not an innocuous tensor coordinate change on the fixed marked section. If the original Killing flow exists only for a short interval, extend the stationary Gaussian-null coefficients to the auxiliary product with $v\in\mathbb R$ and perform the bounded cut change there.  This extension requires no completeness assumption.  The original cut must subsequently be retained as a graph; dropping that graph would change a statement about marked germs into a statement about unmarked germs.

The construction applies equally to nonorientable sections: integrals use densities, and orientation-dependent objects can be lifted to the orientation cover and descended by uniqueness. An actual section is stronger than a compact space of generators. This distinction also appears in the intrinsic rigidity and quasi-Einstein descriptions of horizon geometry \cite{DunajskiLucietti,CollingKatonaLucietti,KaminskiLewandowski2024,Colling2025,CollingDunajskiKunduriLucietti2026}; an explicit NUT obstruction is discussed below.

On either punctured side let $r=\sigma e^{-t}$, $\sigma=\pm1$. Whenever $F>0$, define
\begin{equation}
V=|r|\sqrt F,\quad u=\log V,\quad A=\frac{dr+rh}{r^2F},\quad q=\mu-F^{-1}(dt-h)^2. \label{pos:orbit-definition}
\end{equation}
Then
\begin{equation}
g=e^{2u}(dv+A)^2+q,\qquad q\longrightarrow q_0=\gamma-F_0^{-1}(dt-h_0)^2. \label{pos:kk}
\end{equation}
The convergence and all cylinder derivatives have order $O(e^{-t})$. For the unwarped data, $q_0=-\Lambda^{-1}dt^2+\gamma$ and $du\to-dt$. In general, spacelikeness of $K$ does not by itself provide spacelike $t$-slices. The following criterion provides the uniform Cauchy geometry needed in the proof.

\label{sec:graph-slicing}

The choice $t=-\log|r|$ need not give spacelike slices, even when a different stationary slicing does. The following elementary observation is useful because all the comparison estimates are geometric on the orbit cylinder.

\begin{lemma}[Uniform graph slicing]\label{graph:lemma}
Let $S$ be compact without boundary and let $F,h,\mu$ be smooth Gaussian-null coefficients with $\mu(0)=\gamma$ positive definite. Suppose a single-valued $\psi\in C^\infty(S)$ satisfies
\begin{equation}
\eta(x):=F_0(x)-|h_0+d\psi|_\gamma^2>0\quad(x\in S). \label{graph:condition}
\end{equation}
On either side of $r=0$, set $T=-\log|r|+\psi(x)$. Then on a full late cylinder $[T_1,\infty)\times S$ the orbit metric is uniformly foliated by compact spacelike $T$-slices. Its coefficients and inverse coefficients, the normalized curvature $C=|r|\sqrt F\,dA$, and all their derivatives are bounded in a fixed product metric. The same holds for every positive-order derivative of $u=\log(|r|\sqrt F)$. Equal full Gaussian-null jets give differences decaying faster than $e^{-NT}$ for every $N$, with the same assertion after any fixed number of cylinder derivatives.
\end{lemma}
\begin{proof}
The inverse of the limiting orbit metric is
\[
q_0^{-1}=-F_0\partial_t^2+ \gamma^{ab}(\partial_a+h_{0,a}\partial_t) (\partial_b+h_{0,b}\partial_t).
\]
In particular, the exact identity, including its cross term, is
\begin{equation}
q_0^{-1}(dT,dT)=-F_0+|h_0+d\psi|_\gamma^2. \label{graph:inverse}
\end{equation}
Write $H=h+d\psi$, with all coefficients evaluated at $r=\sigma e^{-T+\psi(x)}$. Direct substitution gives
\begin{equation}
q=\mu-F^{-1}(dT-H)^2,\quad u=-T+\psi+\tfrac12\log F,\quad A=r^{-1}\alpha,\qquad \alpha=F^{-1}(-dT+H). \label{graph:fields}
\end{equation}
The induced slice metric is $\mu-F^{-1}H\otimes H$ and $q^{TT}=-F+|H|_\mu^2$. Compactness supplies $\eta_*:=\min_S\eta>0$; for late $T$, the latter expression is at most $-\eta_*/2$. The slice metric and its inverse are uniformly bounded, as are the lapse $(F-|H|_\mu^2)^{-1/2}$ and shift. This proves the uniform foliation assertion with respect to $dT^2+\gamma$.

Every $T$ derivative acts as $-r\partial_r$, while a spatial derivative at fixed $T$ acts as $\partial_a+r\psi_a\partial_r$. Smoothness on a compact collar therefore gives all the required bounds. In particular
\begin{equation}
du=-dT+d\psi+\tfrac12d\log F,\qquad C=\sigma\sqrt F\{(dT-d\psi)\wedge\alpha+d\alpha\}. \label{graph:normalized-curvature}
\end{equation}
No estimate uses a bound for $u$ itself. Taylor's theorem and boundedness of $\psi$ identify $O(|r|^N)$ with $O(e^{-NT})$ at every fixed differentiated order; division by a fixed power of $r$ preserves flatness. A common late $T$-tail lies in the original collars because $\psi$ is bounded.
\end{proof}

The function $T$ is an auxiliary coordinate on the quotient. It changes neither the horizon cut, the affine Gaussian-null parameter, nor the normalization of $K$. For the limiting metric on $\mathbb R\times S$ it is a temporal function with Cauchy level sets: in its uniformly bounded lapse--shift form every causal curve has bounded spatial speed when parametrized by $T$. A causal curve with a finite limiting $T$ has a spatial limit by this bound and compactness, and is extendible. The same argument applies to finite slabs of the late cylinder.

\begin{proposition}[Geometric meaning and period obstruction]
\label{graph:invariant}
For $F_0>0$, condition \eqref{graph:condition} is equivalent to the existence of a globally spacelike graph section of the stationary orbit cylinder and its translates. Equivalently,
\[
\inf_{\psi\in C^\infty(S)}\ \sup_S\frac{|h_0+d\psi|_\gamma^2}{F_0}<1.
\]
It is also equivalent to stable causality of that cylinder. If it holds, there is a constant $c<1$ such that every piecewise smooth closed curve $\ell$ satisfies
\begin{equation}
\left|\int_\ell h_0\right| \le c\int_\ell\sqrt{F_0}\,|\dot\ell|_\gamma\,ds. \label{graph:period}
\end{equation}
No converse from these period inequalities is used here.
\end{proposition}
\begin{proof}
A graph $t=f(x)$ has induced metric $\gamma-F_0^{-1}(df-h_0)^2$, which is positive definite precisely when $\psi=-f$ satisfies \eqref{graph:condition}. Compactness makes the strict pointwise inequality uniform. The variational formulation follows directly from the definition of an infimum; no minimizer is asserted. A temporal function persists under a sufficiently small widening of the light cones, proving stable causality. Conversely, $\partial_t$ is a complete timelike Killing field and the observer quotient is the smooth compact manifold $S$ with its given trivial product bundle. Harris \cite[Corollary~2.20]{Harris2015} therefore gives a global spacelike section of this same Killing flow whenever $q_0$ is stably causal. That section is a graph, so the first part of the proof supplies $\psi$. This causality equivalence is an application of that established stationary splitting result. Integrating $d\psi$ around $\ell$ gives zero, and the pointwise Cauchy--Schwarz inequality proves \eqref{graph:period}.
\end{proof}

Thus an exact part of $h_0$ can be removed from the criterion, whereas periods of a closed nonexact part may obstruct it. For example, on a flat torus with $F_0=F_*>0$ constant and $h_0=c\,dx^1$, averaging $d\psi$ and applying Jensen's inequality gives $\inf_\psi\sup|h_0+d\psi|^2/F_*=c^2/F_*$. At equality the constant-$t$ $x^1$-circle is null, and above it this circle is timelike. These are tests of the geometric criterion, not vacuum horizon examples.

In vacuum the generalized condition still forces $\Lambda>0$. The intrinsic $vr$ equation is $F_0=\Lambda+\frac12|h_0|^2-\frac12\operatorname{div}_\gamma h_0$. Multiplication by $e^{2\psi}$ and integration by parts give the useful identity
\begin{equation}
\int_S e^{2\psi}\eta\,d\mu_\gamma =\int_S e^{2\psi} \left(\Lambda-\tfrac12|h_0+d\psi|^2 -\tfrac12|d\psi|^2\right)d\mu_\gamma. \label{graph:lambda}
\end{equation}
Its positive left-hand side excludes $\Lambda\le0$. Integration uses densities, so no orientation assumption is necessary.

We write the adapted graph time again as $t$ when no distinction is needed. Relative to $b=dt^2+\gamma$, a tensor is called flat if
\begin{equation}
|(\nabla^b)^k T|_b\le C_{k,N}e^{-Nt} \quad\text{for every }k,N\ge0. \label{pos:flat-cylinder}
\end{equation}
Each individual connection $A$ has at most $e^t$ growth, while differences of connections with identical horizon jets are flat. This elementary separation is used when the fiber is reconstructed.

\section{Unique Continuation}\label{sec:unique-continuation}\label{sec:comparison}\label{strict:section}

The orbit cylinder turns transverse flatness into decay faster than any exponential. What remains is to show that two exact solutions with such a difference must coincide. For a degenerate horizon this conclusion does not follow from the usual nondegenerate horizon extensions \cite{FriedrichRaczWald1999,RaczWald1996,PetersenWave,Petersen,KP}, nor from formal transverse expansions alone \cite{MarsI,MarsII,SPThesis}. The next theorem makes the needed geometric assumption explicit.

\begin{theorem}[Smooth continuation from a spacelike orbit slicing]
\label{strict:theorem}
Let $g_1,g_2$ be smooth $d$-dimensional Einstein metrics, $d\geq4$, with the same Einstein constant $\Lambda$.  Suppose they have degenerate Killing horizons with identified compact connected spacelike sections $S$ without boundary and nonvanishing, normalized generators.  In common marked Gaussian-null coordinates write
\[
g_i=r^2F_i\,dv^2+2\,dv\,dr+2r h_{i,a}\,dv\,dx^a +\mu_{i,ab}\,dx^a dx^b.
\]
Assume the full transverse jets agree and there exists a single global smooth function $\psi:S\to\mathbb R$ such that the common horizon data obey
\begin{equation}
F_0>|h_0+d\psi|_\gamma^2,\qquad \gamma=\mu_i|_{r=0}. \label{strict:sign}
\end{equation}
Then the metrics are smoothly isometric on a two-sided neighborhood of the whole section, preserving the generator and marking.  In the common normalized Gaussian-null coordinates they are equal on a smaller collar. No unwarped or hypersurface-orthogonality assumption is imposed.
\end{theorem}

The sign condition forces $\Lambda>0$ by \eqref{graph:lambda}. The conclusion is local to the section; no global black-hole exterior is assumed.

By Lemma~\ref{graph:lemma}, both sides of the horizon have uniform orbit cylinders. Set $p=du$ and $C=VdA$. Their coefficients and all fixed-order derivatives are bounded, and the differences of $q,u,C,A$ are flat. The exact reduced vacuum equations, with all contractions taken in $q$, are
\begin{align}
dC&=p\wedge C,&\nabla_iC^{ij}&=-2p_iC^{ij},\label{strict:curvature}\\
\operatorname{Ric}_{ij}(q)&=\nabla_i\nabla_j u+p_ip_j+\Lambda q_{ij} +\tfrac12 C_{ik}C_j{}^k,\label{strict:ricci}\\
\square_q u&=-|p|_q^2-\Lambda+\tfrac14C_{ij}C^{ij}. \label{strict:scalar}
\end{align}
Indeed $d(VdA)=p\wedge C$, while the mixed Einstein equation $\nabla_i(V^3(dA)^{ij})=0$ gives the second equation in \eqref{strict:curvature}.  The horizontal and unit-fiber Ricci components give \eqref{strict:ricci}--\eqref{strict:scalar}.

\label{kk:independent-derivation}

The quotient has dimension $d-1$ and the fiber is spacelike.  Write $e^0=V(dv+A)$, $p=d\log V$, $C=VdA$, and use a pseudo-orthonormal quotient coframe $e^a$.  The index $0$ in this paragraph labels the spacelike fiber. The torsion equations $de^A+\omega^A{}_B\wedge e^B=0$ and metric compatibility give
\begin{align*}
\omega^0{}_a&=p_a e^0+\tfrac12 C_{ab}e^b,& \omega^a{}_0&=-p^a e^0-\tfrac12 C^a{}_b e^b,\\
\omega^a{}_b&=\bar\omega^a{}_b-\tfrac12 C^a{}_b e^0.
\end{align*}
Equivalently, for $E_0=V^{-1}K$ and horizontal lifts $E_a$, the Koszul formula gives
\begin{align*}
\nabla^g_{E_a}E_b&=(\nabla^q_aE_b)^{\rm hor}-\tfrac12C_{ab}E_0,& \nabla^g_{E_a}E_0&=\tfrac12C_a{}^bE_b,\\
\nabla^g_{E_0}E_a&=p_aE_0+\tfrac12C_a{}^bE_b,& \nabla^g_{E_0}E_0&=-p^aE_a,
\end{align*}
where $[E_a,E_b]$ has vertical part $-C_{ab}E_0$ and $[E_a,E_0]=-p_aE_0$. For example,
\[
g(R(E_a,E_0)E_0,E_b) =-\nabla_a p_b-p_ap_b+\tfrac14C_{ac}C_b{}^c.
\]
Its quotient trace fixes both the sign and the factor in the fiber equation. Contracting the remaining curvature components, with quotient indices raised by $q$, gives the off-shell identities
\begin{align}
R_{00}(g)&=-\square_q u-|p|_q^2+\tfrac14 C_{ab}C^{ab},\nonumber\\
R_{0a}(g)&=-\tfrac12(\nabla^b C_{ba}+2p^bC_{ba}),\label{kk:ricci-identities}\\
R^{\mathrm{hor}}_{ab}(g)&=\Ric(q)_{ab}-\nabla_a p_b-p_ap_b -\tfrac12 C_{ac}C_b{}^c.\nonumber
\end{align}
These identities are independent of the quotient dimension. In particular $R(g)=R(q)-2V^{-1}\square_qV-\tfrac14V^2(dA)_{ab}(dA)^{ab}$. The sign of the mixed component refers to the horizontal frame $\partial_a-A_a\partial_v$ and the curvature convention used throughout.

For a second derivation, discard a divergence from the Einstein--Hilbert functional with cosmological term $(d-2)\Lambda$.  Per unit Killing coordinate length its reduced density is
\begin{equation}
\sqrt{|q|}\,V\{R(q)-(d-2)\Lambda-\tfrac14 V^2(dA)_{ab}(dA)^{ab}\}. \label{kk:reduced-action}
\end{equation}
The $A$ variation gives $\nabla_a(V^3(dA)^{ab})=0$.  The $V$ variation is $R(q)=(d-2)\Lambda+\tfrac34 C_{ab}C^{ab}$.  The $q$ variation, followed by its trace, gives
\begin{align*}
V^{-1}\square_qV&=-\Lambda+\tfrac14C_{ab}C^{ab},\\
\Ric(q)_{ab}&=V^{-1}\nabla_a\nabla_b V+\Lambda q_{ab} +\tfrac12C_{ac}C_b{}^c.
\end{align*}
Together with $dC=p\wedge C$ these are exactly \eqref{strict:curvature}--\eqref{strict:scalar}.  This variational check is local: neither a complete Killing orbit nor a compact fiber is needed.

\subsection{The terminal gauge and smooth continuation}\label{sec:terminal-gauge}

We record the analytic hypotheses used below. On a cylinder $[a,\infty)\times S$ with compact boundaryless $S$, call a Lorentzian metric $q$ \emph{uniformly foliated} if it can be written
\begin{equation}
q=-\alpha^2\dd t^2+ \gamma_{ab}(\dd x^a+\xi^a\dd t)(\dd x^b+\xi^b\dd t) \label{pos:adm-cylinder}
\end{equation}
with $0<c\leq\alpha\leq C$, the eigenvalues of $\gamma$ uniformly bounded above and below relative to a fixed positive metric on $S$, and the shift, coefficient derivatives, and inverse coefficient derivatives under consideration uniformly bounded. In particular $q^{tt}$ is uniformly bounded away from zero on the negative side.

Consider a section $z$ of a fixed finite-rank bundle with fixed positive bundle metric and fixed uniformly bounded connection $D$, satisfying
\begin{equation}
q^{ij}D_iD_jz=a^iD_i z+b_1z+f. \label{pos:linear-wave}
\end{equation}
At basic order a positive Cauchy energy is equivalent to
\begin{equation}
E_0(t)=\int_{S_t}\left( |\alpha^{-1}(D_t-\xi^aD_a)z|^2+ \gamma^{ab}\langle D_a z,D_bz\rangle+|z|^2\right)\dd\mu_\gamma. \label{pos:positive-energy}
\end{equation}
Here is a direct derivation, including the commutation issue. Put $p=\alpha^{-1}(D_t-\xi^aD_a)z$ and $w_a=D_az$. Up to bounded zeroth-order bundle terms, the equation is the first-order system
\begin{align}
D_tz&=\alpha p+\xi^aw_a,\nonumber\\
D_tw_a&=\xi^bD_bw_a+\alpha D_ap +(D_a\alpha)p+(D_a\xi^b)w_b+\mathcal R_a z,\nonumber\\
D_tp&=\xi^aD_ap+\alpha\gamma^{ab}D_aw_b +L_1(p,w,z)-\alpha f.\label{eq:firstorder-energy}
\end{align}
The coefficients of $L_1$ and $\mathcal R$ are bounded by the first coefficient derivatives, the prescribed connection curvature, and the lower-order coefficients of the original equation. Differentiate $\int(|p|^2+\gamma^{ab}\langle w_a,w_b\rangle+|z|^2)\,d\mu_\gamma$. The two principal spatial terms combine into a divergence; integration by parts leaves only derivatives of $\alpha\gamma^{ab}$ and of the volume form. The shift terms likewise give the spatial divergence of the shift. All these coefficients, and $\partial_t\gamma$, are bounded. Cauchy--Schwarz therefore gives
\[
|\partial_tE_0|\leq C E_0+C\sqrt{E_0}\|f\|_{L^2}.
\]
There is no lateral boundary term. Commute the first-order system, rather than just its scalar second-order expression, with spatial covariant derivatives through order $s$. A commutator with a first-order principal operator has at most $s$ derivatives of $(p,w,z)$; its other factor is a bounded derivative of a known coefficient. Curvature commutators have lower order. Summing the resulting positive energies gives square roots $e_s$ uniformly equivalent to
\begin{equation}
\|z\|_{H^{s+1}(S)}+\|\partial_tz\|_{H^s(S)}. \label{pos:energy-equivalence}
\end{equation}
The resulting backward estimate on every finite slab is
\begin{equation}
e_s(t)\leq e^{c_s(T-t)}e_s(T) +C_s\int_t^T e^{c_s(\tau-t)}\|f(\tau)\|_{H^s}\,\dd\tau. \label{pos:backward-energy}
\end{equation}
Constants depend on uniform finite-order coefficient bounds, not on $T$ or slab length. At zero energy one can regularize the square root or use upper Dini derivatives. The $L^2$ term in \eqref{pos:positive-energy} controls zero modes. Consequently a flat solution of a homogeneous system \eqref{pos:linear-wave} vanishes: take $N>c_s$ in $e_s(T)\leq C_Ne^{-NT}$ and let $T\to\infty$.

For a semilinear fixed-principal-part system with smooth nonlinearity in $(z,Dz)$, vanishing at $(0,0)$, the same estimate holds on a fixed bounded low-Sobolev bootstrap. At higher orders product and composition estimates are tame: the highest norm enters linearly, with coefficients controlled by the low norm and known smooth coefficients. Quadratic products of first derivatives are estimated by placing one factor in $L^\infty$. This is the semilinear local Cauchy theory used in the next lemma; the argument below supplies the required uniformity on long terminal slabs. General wave-gauge Cauchy theory is standard \cite{CB}; smooth Fuchsian energy arguments provide related context \cite{Rendall}, but neither reference is used as a substitute for the gauge construction below.

\label{app:terminal-precise}
We isolate the analytic construction before applying it to a geometric gauge.  Bounds of all orders below mean one finite bound for each order; they do not mean one bound uniform in the order.  Fix a compact $m$-manifold $S$, a smooth metric $\gamma_*$, and $b=dt^2+\gamma_*$.  Write $\mathcal B=\mathbb R\oplus TS$ with its positive $b$-metric and product connection.  The prescribed Lorentzian metrics $q,q_0$ have uniform ADM bounds, and their coefficients and inverses are bounded in $C^j(b)$ for every finite $j$.  The functions $u,u_0$ need not be bounded, but their derivatives of orders $1$ through $j+1$ are bounded for every $j$.  Assume $q-q_0,u-u_0$ are flat in all these norms.

\begin{lemma}[Uniform terminal semilinear lemma]\label{lem:terminal-precise}
Let $s_0>m/2+2$ be an integer.  Suppose on a fixed displacement ball the bundle equation
\[
q^{ij}\nabla_i^b\nabla_j^b Z=f+G(t,x,Z,\nabla^bZ)
\]
has a smooth nonlinearity with $G(t,x,0,0)=0$ whose derivatives are bounded uniformly in $(t,x)$ on bounded sets of $(Z,\nabla^bZ)$, and a flat smooth forcing $f$.  Then there are $t_1$ and $\varepsilon>0$, chosen using only a fixed low-order coefficient class and the forcing, such that for every terminal time $T\geq t_1$ the zero Cauchy problem at $T$ exists on $[t_1,T]$ and obeys
\[
\sup_{t_1\leq t\leq T} e^{Mt}\left(\|Z_T(t)\|_{H^{s+1}} +\|\partial_tZ_T(t)\|_{H^s}\right)\leq C_{s,M}
\]
for every integer $s\geq s_0$ and every $M>0$.  The constants are independent of $T$, and the same $t_1$ works for all $s,M$.  There is a smooth flat limit solving the equation on this common tail.
\end{lemma}
\begin{proof}
Put $P=\alpha^{-1}(\nabla_t^b-\xi^a\nabla_a^b)Z$ and $W_a=\nabla_a^bZ$ in the ADM decomposition of $q$.  The equations for $(Z,P,W)$ form a first-order symmetric hyperbolic system.  A positive symmetrizer is the direct sum of the fixed bundle metric for $Z,P$ and $\gamma^{ab}$ times that metric for $W_a$.  The spatial principal terms are $\xi^a\nabla_a$ and the symmetric $P$--$W$ pair $\alpha\gamma^{ab}\nabla_aW_b,\alpha\nabla_aP$. Commuting spatial derivatives through order $s$, integrating over $S$, and cancelling this pair gives the following inequality after the nonlinear term is absorbed by the Moser estimate proved just below (before absorption the source is $f+G$):
\[
e_s(t)\leq e^{c_s(T-t)}e_s(T) +C_s\int_t^T e^{c_s(\tau-t)}\|f(\tau)\|_{H^s}\,d\tau.
\]
Here $e_s^2$ is the sum of the commuted positive energies, including $\|Z\|_{L^2}^2$.  It is equivalent to $\|Z\|_{H^{s+1}}^2+\|Z_t\|_{H^s}^2$.  A fixed connection which is not metric for the chosen positive bundle metric contributes only bounded zeroth-order terms.  All constants use finitely many coefficient bounds, uniform spacelike-slice bounds and Sobolev constants of $(S,\gamma_*)$; none depend on $T$ or $T-t$.

Here and in the following application the nonlinear estimate is part of the argument, not an additional existence assertion.  On $e_{s_0}\leq\varepsilon$, the values of $(Z,\nabla Z)$ remain in one compact argument set.  For $s\geq s_0$, the spatial Moser product and composition inequalities applied to
\[
G(t,x,Z,P)=\int_0^1 \left(G_Z(t,x,\theta Z,\theta P)Z+ G_P(t,x,\theta Z,\theta P)P\right)d\theta
\]
give $\|G\|_{H^s}\leq C_s e_s$.  The spatial derivatives of the known coefficients are included in $C_s$.  The highest derivative of $\nabla Z$ is controlled by $e_s$; no extra derivative of $Z$ is required. Absorb this linear term in $c_s$ in the preceding estimate.

Choose $N_*>c_{s_0}$ once.  Since $e_s(T)=0$ for every $s$ and $\|f(t)\|_{H^{s_0}}\leq B_{N_*}e^{-N_*t}$,
\[
e_{s_0}(t)\leq \frac{C_{s_0}B_{N_*}}{N_*-c_{s_0}}e^{-N_*t}.
\]
Choose $t_1$ so that this is less than $\varepsilon/2$ for all $t\geq t_1$. Local semilinear Cauchy existence and continuation (see \cite{CB,TaylorPDEIII} for the local theory) in $H^{s_0+1}\times H^{s_0}$ then extend every terminal solution to $t_1$: at a putative finite left endpoint its norm is strictly within the same fixed ball, and a uniform local existence interval contradicts maximality.  Smooth persistence follows by the high-order energy estimate, whose coefficients are controlled by this low norm.  The initial local interval and its continuation do not require smallness of any high norm.

For a prescribed pair $(s,M)$ choose $N>\max(c_s,M)$.  The same estimate, now using $\|f(t)\|_{H^s}\leq B_{s,N}e^{-Nt}$, yields the asserted decay with no change of $t_1$.  The constants $C_{s,M}$ may increase with $s,M$; this affects neither existence nor the common tail.  Uniform negativity of $q^{tt}$ permits solving for $Z_{tt}$.  Differentiating the equation recursively gives the corresponding estimates for every mixed time and space derivative.  On each compact slab, the all-order Sobolev bounds, Sobolev embedding and Arzel\`a--Ascoli give a convergent subsequence in every $C^k$ norm.  A diagonal subsequence yields a smooth solution on the whole tail.  Its estimates follow by passage to the limit.  Equivalently, a difference estimate on $[t_1,\min(T,T')]$ proves uniqueness of this flat limit, using its rapidly decaying data at the smaller terminal time.
\end{proof}

\begin{lemma}[Terminal weighted wave-map gauge]\label{pos:terminal-gauge}
Let $q,q_0$ be smooth uniformly foliated Lorentzian metrics on a cylinder $[a,\infty)\times S$, with compact boundaryless $S$. Assume bounded coefficients, inverses and derivatives of each order. Let $u,u_0$ have bounded positive-order derivatives and assume $q-q_0,u-u_0$ are flat. Set $W=\operatorname{grad}_{q_0}u_0$. On one late tail there is a smooth flat perturbation $Y$ of the identity satisfying
\begin{equation}
\tau_{q,q_0}(Y)+dY(\operatorname{grad}_q u)-W\circ Y=0. \label{pos:weighted-wave-map}
\end{equation}
It is a diffeomorphism onto a region containing a complete later tail; its inverse is also flat relative to the identity.
\end{lemma}
\begin{proof}
Fix $b=dt^2+\gamma_*$ and write $Y=\operatorname{Exp}^b_{(t,x)}Z$, with $Z\in\mathbb R\oplus TS$. The compact section has a uniform injectivity radius. In the chosen small displacement ball the displacement differential $d_Z\operatorname{Exp}^b$ is invertible. In local coordinates,
\[
\tau(Y)^\alpha=q^{ij}\bigl(\partial_{ij}Y^\alpha -\Gamma(q)^k{}_{ij}\partial_kY^\alpha +\Gamma(q_0)^\alpha{}_{\beta\gamma}(Y) \partial_iY^\beta\partial_jY^\gamma\bigr).
\]
The only unknown second-derivative term in $\partial_{ij}Y$ is $(d_Z\operatorname{Exp}^b)\partial_{ij}Z$. Multiplying by its inverse therefore leaves exactly $q^{ij}\nabla_i^b\nabla_j^b Z$ as principal part. Every remaining term contains at most first derivatives of $Z$. With
\[
\mathcal C(q,q_0)^k=q^{ij}(\Gamma(q)^k{}_{ij}-\Gamma(q_0)^k{}_{ij}),
\]
the defect of the identity in the geometric equation is $-\mathcal C(q,q_0)+\operatorname{grad}_q u-W$, which is flat. Thus the displacement equation has precisely the hypotheses of Lemma~\ref{lem:terminal-precise}; its forcing is the negative identity defect after the invertible displacement differential is applied. The Lorentzian target signature appears only in known lower-order coefficients. The positive bundle metric is $b$. Only derivatives of $u,u_0$ occur, and their unbounded values cause no coefficient growth.

The lemma constructs a smooth flat solution on a common tail. For each $t$, the spatial map is a local diffeomorphism of $S$ homotopic to the identity. By compactness it is a covering, and its induced fundamental-group map is an isomorphism, so the covering has one sheet. This argument requires no orientation. Hold the target spatial point fixed and invert that spatial map. The target time then has derivative $1+O(\varepsilon)>0$, proving global injectivity and that the image contains a complete later product tail. Repeated implicit differentiation gives flat bounds for the inverse. Starting the source a fixed distance later avoids the left-end coordinate issue.
\end{proof}

\begin{proof}[Proof of Theorem~\ref{strict:theorem}]
Apply Lemma~\ref{pos:terminal-gauge} with the general reference pair $(q_2,u_2)$ and $W=\operatorname{grad}_{q_2}u_2$.  Push the first solution forward by its flat map and suppress the resulting tildes.  The achieved gauge is $\mathcal C(q_1,q_2)=\operatorname{grad}_{q_1}u_1-W$. Consequently $\operatorname{Ric}-\frac12\mathcal L_{\mathcal C}q$ contains no Hessian of $u$ on its right-hand side. At the same time, $\square_q u=q^{ij}D_iD_j u-\mathcal C(u)$, so $\mathcal C(u)=|du|_q^2-W(u)$ cancels the quadratic gradient term in \eqref{strict:scalar}. The scalar equation becomes
\[
q^{ij}D_iD_j u+W(u)=-\Lambda+\tfrac14C_{ij}C^{ij}.
\]
The metric equation has principal part $-\frac12q^{ij}D_iD_jq$ and algebraic curvature terms $C_{ik}C_j{}^k$. For $z=(q_1-q_2,u_1-u_2)$ and $U=C_1-C_2$, subtraction therefore gives
\begin{equation}
q_1^{ij}D_iD_jz=\mathsf A^iD_i z+\mathsf Bz+\mathsf DU, \label{strict:wave-difference}
\end{equation}
with bounded coefficients.  A principal-coefficient difference multiplies known second derivatives of the reference solution and is a zeroth-order term.  Here $\mathsf D$ is a bounded algebraic coefficient map, not a differential operator. No derivative of $U$ occurs.

The curvature differences satisfy, writing $p_i^{\sharp_i}$ for the $q_i$-dual vector,
\begin{align}
dU&=p_1\wedge U+(p_1-p_2)\wedge C_2,\nonumber\\
\nabla^{(1)i}U_{ij} &=-2(p_1^{\sharp_1})^iU_{ij} -2(p_1^{\sharp_1}-p_2^{\sharp_2})^i(C_2)_{ij} +(\nabla^{(2)i}-\nabla^{(1)i})(C_2)_{ij}. \label{strict:form-difference}
\end{align}
The last term is a bounded linear expression in $z,Dz$, using the known first derivatives of $C_2$.  Thus the curl and divergence sources involve only $z,Dz,U$; they introduce no derivative loss.

For the two-form energy use
\[
T_{ij}[U]=U_{ik}U_j{}^k-\tfrac14(q_1)_{ij}U_{kl}U^{kl}, \qquad \nabla^iT_{ij}=(\nabla^iU_{ik})U_j{}^k +\tfrac12U^{ik}(dU)_{ijk}.
\]
If $n$ is the unit timelike normal, then $T(n,n)=\frac12\sum_a U(n,e_a)^2+\frac12\sum_{a<b}U(e_a,e_b)^2$ in a quotient orthonormal frame. Thus its normal energy is positive and uniformly equivalent to $\|U\|_{L^2(S_t)}^2$ in every quotient dimension under consideration. Let $E_U$ be this energy and let $E_z$ be the positive wave energy of \eqref{strict:wave-difference}, including its $L^2$ term as in \eqref{pos:backward-energy}.  Bounded lapse, shift and deformation tensor, \eqref{strict:form-difference}, and Cauchy--Schwarz give
\[
|E_U'|\leq c(E_U+E_z),\qquad |E_z'|\leq c(E_z+E_U).
\]
In slab form the two inequalities say $|E(t_2)-E(t_1)|\le c\int_{t_1}^{t_2}E(s)\,ds$. The source terms are bounded by $|U|(|z|+|Dz|+|U|)$ and the deformation term by $|T[U]|\,|\nabla n|$. There is no lateral boundary because $S$ is compact without boundary. The constant depends on finite coefficient bounds on the fixed tail, not the terminal time.  Hence
\[
E(t)\leq e^{c_0(T-t)}E(T),\qquad E=E_U+E_z.
\]
All differences remain flat after the gauge change, so $E(T)\leq C_Ne^{-2NT}$ for every $N$.  Taking $2N>c_0$ and $T\to\infty$ yields $z=U=0$.

Returning to the original quotient coordinates gives $q_1=Y^*q_2$, $u_1=Y^*u_2$, and $C_1=Y^*C_2$.  Therefore $dA_1=Y^*dA_2$.  The form $B=A_1-Y^*A_2$ is closed and flat: every derivative of $A_2$ grows at most exponentially, while $Y$ is flat relative to the identity. It has the global flat primitive
\[
f(t,x)=-\int_t^\infty B_t(s,x)\,ds, \qquad df=B,
\]
where closedness and $B_a\to0$ give the spatial derivative identity. The lifted map $(v,t,x)\mapsto(v+f(t,x),Y(t,x))$ is an isometry. For $Y=(t+Z^0,x')$ in graph time, its signed radial component is
\[
r'=r\exp\{-Z^0+\psi(x')-\psi(x)\}.
\]
Every fixed $r$-derivative costs only a fixed power of $r^{-1}$. Flatness at every exponential order consequently makes all departures from the identity smooth and flat in signed $r$. The two maps glue across the horizon with identity differential, giving a whole-section two-sided collar by compactness.  Finally, uniqueness of the marked transverse affine null geodesics identifies the common Gaussian-null coordinates and proves their coefficient equality.
\end{proof}

\begin{corollary}[Unwarped full-jet determination]\label{thm:jets}
Let $d\ge4$ and let two smooth Einstein germs have the same $\Lambda>0$, identified compact connected boundaryless actual horizon sections, nonvanishing normalized degenerate Killing generators, and the unwarped data \eqref{eq:unwarped}. Equality of all transverse jets in their common marked Gaussian-null constructions implies a smooth isometry on a uniform two-sided neighborhood of the whole section. It preserves the generator, marked horizon points and transverse null normals. The common normalized coefficient functions are equal.
\end{corollary}
\begin{proof}
Take $\psi=0$ in Theorem~\ref{strict:theorem}.
\end{proof}

\subsection{Finite weighted matching}\label{sec:finite-graph}
The continuation argument itself gives a finite, though implicit, matching order for a general rotating horizon.  This is different from proving that the intrinsic horizon geometry determines those jets.

\begin{theorem}[Finite weighted matching for a fixed orbit cylinder]
\label{thm:finite-graph}
Let $d\geq4$ and let $S$ be a compact connected smooth manifold without boundary, of dimension $d-2$, identified with an actual horizon section. Fix smooth marked horizon data $(\gamma,F_0,h_0)$ on $S$, a nonvanishing normalized degenerate Killing generator, an Einstein constant $\Lambda$, and a smooth function $\psi$ such that $F_0>|h_0+d\psi|_\gamma^2$ everywhere. There is an integer $N$ depending only on these data and on fixed finite coefficient bounds around their limiting orbit cylinder with the following property. If two smooth Einstein germs with these data have matching transverse packages
\[
\left(\partial_r^j\mu, \partial_r^{j+1}(rh), \partial_r^{j+2}(r^2F)\right)\big|_{r=0}, \qquad 1\leq j\leq N,
\]
in their common marked Gaussian-null constructions, they are smoothly isometric on a two-sided neighborhood of the whole section, preserving its marking and normalized generator.  The collar size may depend on the germs.  No unwarped, twist-free or symmetry assumption is imposed.
\end{theorem}
\begin{proof}
Work on either side with $T=-\log|r|+\psi(x)$; use the larger constants of the two sides.  The limiting quotient metric, logarithmic-lapse gradient and normalized curvature are determined by the prescribed horizon data. Fix $s=\lfloor m/2\rfloor+6$, a product metric $dT^2+\gamma$, uniform ADM bounds, a finite $C^{s+4}$ neighborhood of the limiting quotient fields, and a small $C^4$ displacement neighborhood.  Reserve also a fixed low-order coefficient class containing the pushforwards of all fields in the first class by maps in the displacement neighborhood.  All choices are made before choosing either germ.  Denote by $c_G$ the square-root energy growth constant for the terminal gauge in $H^{s+1}\times H^s$, and by $c_E$ the growth constant for the squared, coupled metric--scalar and two-form energy in the reserved class.  These are finite constants of the fixed classes.  Choose an integer $N$ satisfying
\begin{equation}
N>3,\qquad N+1>c_G,\qquad 2(N+1)>c_E. \label{eq:finite-graph-N}
\end{equation}

Every smooth germ with the prescribed data enters the fixed input class on a sufficiently late tail.  Equality of the displayed packages is exactly equality of the derivatives of $F,h,\mu$ through order $N$. Taylor's theorem, including mixed derivatives of any prescribed finite order, gives, with $\nu=N+1$,
\[
q_1-q_2=O(e^{-\nu T}),\quad u_1-u_2=O(e^{-\nu T}),\quad C_1-C_2=O(e^{-\nu T}),\quad A_1-A_2=O(e^{-(\nu-1)T}).
\]
The constants in these remainder estimates may depend on the germs; they will only change the start of the tail.  The identity defect in the gauge equation has $H^s$ size at most $B e^{-\nu T}$.  Solve its zero terminal Cauchy problem at a finite time $L$.  The same low-norm bootstrap as in Lemma~\ref{pos:terminal-gauge} gives
\[
e_s(T)\leq\frac{CB}{\nu-c_G}e^{-\nu T},\qquad t_1\leq T\leq L,
\]
where a single sufficiently late $t_1$ works for all $L$.  Smooth persistence for each finite $L$ follows from the tame high-order estimates; those higher norms are not required to be bounded uniformly as $L\to\infty$.

The fixed estimate at order $s$, the equation solved for $Z_{TT}$, and its first two time derivatives give uniform bounds for
\[
Z,\ Z_T,\ Z_{TT},\ Z_{TTT},\ Z_{TTTT} \quad\hbox{in}\quad H^{s+1},\ H^s,\ H^{s-1},\ H^{s-2},\ H^{s-3},
\]
respectively, all at rate $e^{-\nu T}$.  They imply compact convergence in $C^3$ on finite subcylinders and a $C^3$ limiting gauge map $Y$, with the corresponding finite-rate bounds.  Its inverse is $C^3$ and has the same bounds.  The pushed metric is $C^2$, the pushed normalized two-form is $C^2$, and the achieved gauge and reduced equations hold classically. After increasing $t_1$, their coefficients lie in the low-order class reserved before \eqref{eq:finite-graph-N}.

Set $z=(\widetilde q_1-q_2,\widetilde u_1-u_2)$ and $U=\widetilde C_1-C_2$.  The wave and curl--div systems of the main comparison proof give
\[
E(T)\leq e^{c_E(L-T)}E(L),\qquad E\asymp\|z\|_{H^1}^2+\|\partial_Tz\|_{L^2}^2+\|U\|_{L^2}^2.
\]
The gauge bounds imply $E(L)=O(e^{-2\nu L})$.  Since $2\nu>c_E$, taking $L\to\infty$ gives $z=U=0$.  Consequently $B=A_1-Y^*A_2$ is closed and has rate $e^{-(\nu-1)T}$, together with two cylinder derivatives.  Its global primitive $f=-\int_T^\infty B_T(s,x)\,ds$ satisfies $df=B$ and the same rate. The lifted map is an isometry on the punctured collar.

If $Y=(T+Z^0,x')$, its original radial component is
\[
r'=r\exp\{-Z^0+\psi(x')-\psi(x)\}.
\]
Thus $r'-r$ has order $r^{N+2}$, the angular displacement has order $r^{N+1}$ and the fiber displacement has order $r^N$, including the mixed derivatives needed here.  Since $N>3$, the two side maps extend and glue as a $C^2$ map whose differential is the identity on the horizon. They form a local $C^2$ isometry there.  The connection transformation identity for the two smooth metrics successively improves this map to $C^\infty$.  Finally affine-null-geodesic uniqueness makes it the identity in the two common marked Gaussian-null gauges.
\end{proof}

The integer $N$ is uniform for a family of prescribed horizon data having uniform graph-slicing margin, finite coefficient bounds and Sobolev constants, for example a compact parameter subset of a fixed regular rotating-Nariai family on $S^2$.  Nothing in this statement supplies an optimal or numerically evaluated $N$, permits a margin tending to zero, or classifies the admissible rotating transverse packages.  In the unwarped nonresonant problem, the separate elliptic recursion lowers the matching requirement to the first package.

\begin{remark}[Conditional quantitative comparison]
The inhomogeneous estimate \eqref{pos:backward-energy} controls an earlier slice by full Cauchy data on a later slice and the equation error on the intervening slab, within the prescribed uniform coefficient class. It does not replace those norms by an unspecified Taylor remainder or provide uniform inverse stability for arbitrarily rough angular data.
\end{remark}

\label{unwarped:section}

\subsection{Staticity of unwarped horizons}

The quotient twist method has an antecedent in the smooth staticity theorem of Chru\'sciel--Mars for bifurcate horizons \cite{CMStaticity}. For compact degenerate horizons, the same curvature system gives the following result without a sectional-curvature assumption.

\begin{theorem}
\label{thm:frobenius}
Let $K$ be a nonvanishing smooth Killing generator of a smooth degenerate horizon with an actual compact connected spacelike section $S$ without boundary. Suppose the stationary Gaussian-null data satisfy $h|_S=0$ and $F|_S>0$, and
\[
\operatorname{Ric}_g(K,\cdot)=\lambda K^\flat
\]
on a collar, for a scalar $\lambda$. Then $F|_S$ is constant and $K$ is hypersurface orthogonal on a two-sided neighborhood of the whole section. No Einstein or curvature condition on the section is required. In vacuum, the hypotheses hold for unwarped data $F|_S=\Lambda>0$, and the intrinsic equation supplies $\operatorname{Ric}(\gamma)=\Lambda\gamma$.
\end{theorem}

\begin{proof}
First, constancy of $F_0=F|_S$ follows from the mixed Ricci condition. Under the near-horizon scaling the metric tends to $2\,dv\,dr+r^2F_0(x)\,dv^2+\gamma$. Indeed, under $v\mapsto v/\varepsilon$, $r\mapsto\varepsilon r$, the mixed coefficient is $2r h(\varepsilon r)=O(\varepsilon)$ because $h|_S=0$; the transverse derivative of $h$ does not survive the limit. For the limiting metric, direct computation gives $\operatorname{Ric}_{vr}=F_0$ and $\operatorname{Ric}_{va}=r\partial_aF_0$. The proportionality factor has a smooth limit because it equals $\operatorname{Ric}(K,\partial_r)$ in Gaussian-null coordinates. The limiting eigenvector condition has zero $a$ component on its right-hand side, hence $d_SF_0=0$. Connectedness makes $F_0=a>0$ constant. No other Einstein equation is used.

Use a Gaussian-null collar
\[
g=r^2F\,\dd v^2+2\,\dd v\,\dd r +2r h_a\,\dd x^a\,\dd v+\mu_{ab}\,\dd x^a\,\dd x^b,\qquad K=\partial_v.
\]
On each side put $r=\sigma e^{-t}$, $\sigma=\pm1$, and
\[
V=|r|\sqrt F,\quad u=\log V,\quad A=\frac{\dd r+r h}{r^2F},\quad q=\mu-\frac{(\dd r+r h)^2}{r^2F}.
\]
Then $g=V^2(\dd v+A)^2+q$, and $q$ is Lorentzian. Writing $F=a+rF_1+O(r^2)$, $h=r h_1+O(r^2)$ and $\mu=\gamma+O(r)$ gives
\[
q=-a^{-1}\dd t^2+\gamma+O(e^{-t}),\qquad \dd u=-\dd t+O(e^{-t}),
\]
with the derivatives used below. The mixed Ricci condition gives the established quotient twist system
\begin{equation}
\dd\Omega=0,\qquad \nabla_i^q(V^3\Omega^{ij})=0,\qquad \Omega=\dd A. \label{unwarped:maxwell}
\end{equation}
This is the corresponding normalization of \cite[Sec.~2]{CMStaticity}.

The connection and its curvature have expansions
\[
A_t=-\frac1{a r}+\frac{F_1}{a^2}+O(r),\qquad A_S=\frac{h_1}{a}+O(r),\qquad \Omega=\dd t\wedge E_0+B_0+O(r),
\]
where
\begin{equation}
E_0=-a^{-2}\dd_S F_1,\qquad B_0=a^{-1}\dd_S h_1,\qquad \partial_t\Omega=O(r). \label{unwarped:boundary}
\end{equation}
The coefficients $F_1,h_1$ are global on $S$ because there is a global Gaussian-null collar of the compact section. Thus both limiting forms are exact; no vanishing-cohomology assumption is being made.

Taking the temporal and spatial components of $\nabla_i\Omega^{ij}=-3u_i\Omega^{ij}$ at $t=\infty$ gives
\[
\operatorname{div}_{\gamma}E_0=0,\qquad \operatorname{div}_{\gamma}B_0+3a E_0=0.
\]
Exactness and compactness make $E_0=0$. Hence $B_0$ is co-closed, and integration by parts gives $\|B_0\|_{L^2}^2 =\langle \dd(h_1/a),B_0\rangle =\langle h_1/a,\delta B_0\rangle=0$. These integrations also work with densities or on an orientation cover. Consequently $\Omega=O(e^{-t})$.

It remains to exclude a decaying, nonzero bulk twist. Let $n$ be the future unit normal to the slices, $\alpha$ the lapse, and let
\[
T_{ab}=\Omega_{ac}\Omega_b{}^c-\tfrac14q_{ab}\Omega_{cd}\Omega^{cd}, \qquad \mathcal E(t)=\int_{S_t}T(n,n)\,\dd\mu_{\gamma(t)}.
\]
This energy is positive and controls every component of $\Omega$. For $U=\operatorname{grad}_q u$, equation \eqref{unwarped:maxwell} gives
\begin{equation}
\nabla^aT_{ab}=-3U^a\Omega_a{}^c\Omega_{bc}. \label{unwarped:divergence}
\end{equation}
Write $U=c n+U_S$, with $c=-n(u)$ and $U_S$ tangential. The cylinder asymptotics imply $\alpha c=1+O(e^{-t})$, $U_S=O(e^{-t})$, and $\nabla n=O(e^{-t})$. Apply the divergence theorem to $P^a=T^a{}_b n^b$ on compact slices. Since $\alpha\sqrt{\det\gamma}\,P^t=-\sqrt{\det\gamma}\,T(n,n)$, \eqref{unwarped:divergence} yields
\begin{equation}
\mathcal E'(t)=3\int_{S_t}\alpha c\,|i_n\Omega|^2\,\dd\mu_{\gamma(t)} +R(t),\qquad |R(t)|\leq Ce^{-t}\mathcal E(t). \label{unwarped:energy}
\end{equation}
The first term is nonnegative on a late tail. Thus, for $T>t$,
\[
\mathcal E(t)\leq \exp\!\bigl(C(e^{-t}-e^{-T})\bigr)\mathcal E(T).
\]
Since $\mathcal E(T)=O(e^{-2T})$, letting $T\to\infty$ gives $\Omega=0$.

This argument applies on both sides, since $\dd r=-r\,\dd t$ on each. Finally,
\[
K^\flat\wedge\dd K^\flat=V^4(\dd v+A)\wedge\Omega.
\]
It vanishes on both punctured collars, and hence on the horizon by continuity. Near $S$ the field $K$ is nonzero, so Frobenius gives the asserted local hypersurface orthogonality.
\end{proof}

In the vacuum specialization the intrinsic horizon equation supplies $\Ric(\gamma)=\Lambda\gamma$, but no classification of that Einstein metric or of the remaining spacetime is asserted. The hypothesis is the unwarped near-horizon product, not an arbitrary warped static positive-$\Lambda$ near-horizon geometry. The global compact section is essential: without it, a nontrivial circle bundle can carry nonexact limiting magnetic curvature, and the integration step in \eqref{unwarped:boundary} is unavailable. Here hypersurface orthogonality does not mean that $K$ is timelike; it is spacelike on both punctured sides. No global static exterior or causal admissibility is concluded.

The energy argument uses only finitely many derivatives. In a prescribed stationary Gaussian-null collar, $C^3$ regularity of $(F,h,\mu)$ suffices: the divided smooth coefficients and their first two derivatives have the limits used in the constraint step, and the current identity uses only one derivative of $q$ and of the normal. This is a sufficient coordinate-regularity statement; no optimal regularity theorem for a rough horizon or rough Killing generator is asserted.

\begin{remark}[NUT sharpness of the whole-section hypothesis]
The local Taub--NUT--de Sitter family \cite{NUTdS} makes the topological restriction concrete. Put $D=\rho^2+b^2$, $\Theta=\dd v+2b\cos\vartheta\,\dd\varphi$, and
\[
g=-f\Theta^2+2\,\dd\rho\,\Theta+D\,\dd\Omega_2^2,\qquad f=\frac{\rho^2-2M\rho-b^2-(\Lambda/3)(\rho^4+6b^2\rho^2-3b^4)}{D}.
\]
Choose real $(\rho_0,b)$ with $b\ne0$ and $\rho_0^2+b^2=\Lambda^{-1}$, and put $M=\rho_0(1-4\Lambda b^2)/3$. Then $f(\rho_0)=f'(\rho_0)=0$ and $f''(\rho_0)=-2\Lambda$. The regular angular-chart metric has unwarped near-horizon limit $2\dd v\dd r+\Lambda r^2\dd v^2+\Lambda^{-1}\dd\Omega_2^2$, yet $\dd\Theta=-2b\sin\vartheta\,\dd\vartheta\wedge\dd\varphi\ne0$ and the ambient field is not hypersurface orthogonal. Its magnetic flux is $-8\pi b$. Removing the polar gauge strings requires the nontrivial NUT circle bundle, which admits no global spacelike section of its null generators. A compact transverse multisection would project as a finite cover and retain nonzero flux, giving the same obstruction. Thus this is a local-patch counterexample to dropping the whole compact section, not a counterexample to Theorem~\ref{thm:frobenius}. Neither the family nor its connection curvature is claimed as new.
\end{remark}

\section{Transverse Rigidity}\label{sec:transverse-rigidity}
\label{nr:sec}

\subsection{Transverse recursion and spectral conditions}

Unique continuation becomes a classification theorem once the horizon equations determine the transverse derivatives. For an unwarped section the relevant linearization acts on the entire symmetric-tensor bundle. Its trace and divergence sectors cannot be discarded at the outset, even though the final difference equation is trace-free. The first-order deformation problem goes back to \cite{LiLucietti2016,KL,FontanellaGutowski2017}; here we use the full operator and then pass from the formal recursion to actual smooth metrics.

Let $(S^m,\gamma)$ be compact and connected, without boundary, with $\operatorname{Ric}(\gamma)=\Lambda\gamma$, $\Lambda>0$. The curvature convention throughout this section is
\[
(\mathring R k)_{ab}=R_{acbd}k^{cd},\qquad \Delta_Lk=\nabla^*\nabla k+2\Lambda k-2\mathring R k.
\]
Thus $\mathring R\gamma=\Lambda\gamma$ and $\Delta_L\gamma=0$. Our convention agrees with the usual geometric Lichnerowicz operator \cite{Besse1987}; Katona--Lucietti \cite{KL} use one half of this $\Delta_L$ in their equation (25).

Use stationary Gaussian-null coordinates
\[
g=2\,dv\,dr+\phi\,dv^2+2\beta_a\,dv\,dx^a+\mu_{ab}\,dx^a dx^b, \quad \phi=r^2F,\quad\beta=rh,
\]
with $F(0)=\Lambda$, $h(0)=0$, and $\mu(0)=\gamma$. A superscript $(j)$ denotes the $j$th $r$ derivative at zero, without absorbing a factorial. For an actual cut $v=f(x)$, the first-jet supertranslation is $\mu^{(1)}\mapsto\mu^{(1)}-2\nabla^2 f$ when $h_0=0$. The parameter in Katona--Lucietti's formula $\mu^{(1)}\mapsto\mu^{(1)}+\nabla^2 f_{\rm KL}$ is $f_{\rm KL}=-2f$. Indeed the new null normal is $N_f=N-\nabla f-\tfrac12|df|^2K$; evaluating its tangential second fundamental form gives the displayed Hessian term when $h_0=0$. Solving the scalar Poisson equation fixes $\operatorname{tr}_\gamma\mu^{(1)}$ to be constant. This is a change of the cut, so an arbitrary original marked cut must subsequently be retained as its graph in the normalized collar.

\begin{lemma}[First-jet equation]
\label{nr:first}
In constant-trace gauge the first transverse coefficient $k=\mu^{(1)}$ satisfies
\[
\Delta_Lk=0,\qquad \beta^{(2)}=-\operatorname{div}k,\qquad \phi^{(3)}=-\Lambda\operatorname{tr}k-\operatorname{div}\beta^{(2)}.
\]
These formulas require the Einstein property of $\gamma$, but not constant sectional curvature.
\end{lemma}
\begin{proof}
The $ra$ equation at order zero and $vr$ equation at order one give the last two formulas. The $ab$ equation at order one gives
\[
(D\operatorname{Ric})_\gamma(k)_{ab} =\nabla_{(a}\nabla^c k_{b)c}.
\]
For an Einstein background, the variation formula is
\[
(D\operatorname{Ric})_\gamma(k)_{ab} =\tfrac12(\Delta_Lk)_{ab} +\nabla_{(a}\left(\nabla^c k_{b)c} -\tfrac12\nabla_{b)}\operatorname{tr}k\right).
\]
The constant trace removes the final Hessian and the divergence terms cancel. These are the equations preceding the constant-curvature specialization in Katona--Lucietti (21)--(26).
\end{proof}

\begin{lemma}[Leading difference]
\label{nr:difference}
Suppose two stationary Einstein metrics have the same normalized near-horizon data and the same transverse packages through order $n-1$, where $n\geq2$. Put
\[
k=\mu_1^{(n)}-\mu_2^{(n)},\quad b=\beta_1^{(n+1)}-\beta_2^{(n+1)},\quad p=\phi_1^{(n+2)}-\phi_2^{(n+2)}.
\]
Then
\[
\operatorname{tr}k=0,\qquad b=-\operatorname{div}k, \qquad p=-\operatorname{div}b,
\]
and
\begin{equation}
\mathcal L_n k=0,\qquad \mathcal L_n:=\Delta_L+[n(n+1)-2]\Lambda =\nabla^*\nabla-2\mathring R+n(n+1)\Lambda. \label{nr:indicial}
\end{equation}
\end{lemma}
\begin{proof}
Subtract the $rr$ equation at derivative order $n-2$. Its highest new coefficient is $-\tfrac12\operatorname{tr}k$; all other terms involve lower matched packages. This gives the trace constraint. The highest new terms in the $ra$ equation at derivative order $n-1$ are $\tfrac12[b+\operatorname{div}k-d\operatorname{tr}k]$, hence the formula for $b$. The $vr$ equation at order $n$ gives $p=-\operatorname{div}b$, since its remaining highest-order scalar term is proportional to $\operatorname{tr}k$.

In the $ab$ equation at derivative order $n$, the new terms are
\[
(D\operatorname{Ric})_\gamma(k) +\nabla_{(a}b_{b)}-\Lambda k +\tfrac12n(n+1)\Lambda k=0.
\]
The coefficient $n(n+1)$ follows directly from $\partial_r(r^2\partial_r r^n)=n(n+1)r^n$. The variation formula from Lemma~\ref{nr:first}, the trace constraint, and $b=-\operatorname{div}k$ cancel every divergence term. Multiplying by two gives \eqref{nr:indicial}. Products containing a new coefficient and a positive-order background coefficient have strictly higher transverse weight; hence this argument compares arbitrary backgrounds sharing the lower jets, not only a product model.
\end{proof}

The operator in \eqref{nr:indicial} acts on all trace-free symmetric 2-tensors; no transverse-traceless restriction was assumed. It is elliptic and self-adjoint on the compact section. Here is a precise inhomogeneous recursion, including its factorial normalization. Form the lower polynomial $\bar g_{<n}$ by retaining $\mu^{(j)}$ through $j=n-1$, $\beta^{(j+1)}$ through $j=n-1$, and $\phi^{(j+2)}$ through $j=n-1$, with the Taylor factorials $j!$, $(j+1)!$ and $(j+2)!$, respectively. The base polynomial is $\mu=\gamma$, $\beta=0$, $\phi=\Lambda r^2$; in raw derivative notation $\phi^{(2)}=2\Lambda$, not $\Lambda$. For $\mathcal E(g)=\operatorname{Ric}(g)-\Lambda g$ define the four known residuals
\begin{align}
T_n&=\partial_r^{n-2}\mathcal E_{rr}(\bar g_{<n})|_0,& (B_n)_a&=\partial_r^{n-1}\mathcal E_{ra}(\bar g_{<n})|_0,\nonumber\\
P_n&=\partial_r^n\mathcal E_{vr}(\bar g_{<n})|_0,& (Q_n)_{ab}&=\partial_r^n\mathcal E_{ab}(\bar g_{<n})|_0. \label{nr:residuals}
\end{align}
For the new coefficients $k,b,p$, put $\tau=2T_n$ and write $k=k_0+(\tau/m)\gamma$. The equations are exactly
\begin{align}
\operatorname{tr}k&=\tau,& b&=-\operatorname{div}k+d\tau-2B_n,\nonumber\\
p&=-\operatorname{div}b-n\Lambda\tau-2P_n,& \mathcal L_n k_0&= \left[-2Q_n-\nabla^2\tau+4\nabla_{(a}(B_n)_{b)}\right]_0. \label{nr:source}
\end{align}
The final subscript means trace-free projection. The trace of the unprojected angular equation remains a compatibility condition; so do any unused components of the Einstein residual. For existing Einstein germs they hold automatically. We do not infer formal existence for arbitrary lower data from this elimination alone. The range of $\mathcal L_n$ is the $L^2$ orthogonal complement of its finite-dimensional kernel; the displayed source must be orthogonal to that kernel. On its orthogonal complement elliptic regularity gives $\|k_0\|_{H^{s+2}}\le C_s\|\mathcal L_nk_0\|_{H^s}$. Thus both the obstruction and the reconstruction are checkable from the lower jets. The constant-trace cut has fixed the relevant first-order supertranslation freedom. No independent higher-order Gaussian-null gauge transformation is discarded in Lemma~\ref{nr:difference}.

For completeness, the scalar- and vector-derived sectors introduce no additional negative-spectrum resonance when $\Lambda>0$. On an Einstein manifold, $\Delta_L(f\gamma)=(-\Delta f)\gamma$ and $\Delta_L\operatorname{sym}\nabla\alpha =\operatorname{sym}\nabla(\Delta_H\alpha)$. The analogous identity holds for the trace-free Hessian of a scalar eigenfunction. Therefore the nonconstant scalar and Hodge-derived indicial denominators are their nonnegative eigenvalues plus $[n(n+1)-2]\Lambda$; for $n\geq2$ these are strictly positive. At $n=1$, positive Ricci curvature excludes harmonic one-forms, and the remaining trace-free kernel is transverse-traceless. A genuine order-$n$ TT resonance has the precise condition
\[
\lambda_L=[2-n(n+1)]\Lambda, \qquad \lambda_L\in\operatorname{spec}(\Delta_L|_{\mathrm{TT}}).
\]
The scalar, vector and TT decomposition is not needed for the coercivity argument below; using the full operator is stronger and avoids a possible omitted sector.

\begin{theorem}[Finite-jet determination of smooth germs]
\label{thm:finite-explicit}
For a compact positive Einstein section let
\begin{align}
\kappa_\gamma&=\max_{x\in S}\lambda_{\max} \left(\mathring R_x|_{\operatorname{Sym}^2_0T_x^*S}\right), &\kappa_+&=\max(0,\kappa_\gamma),\nonumber\\
J_\gamma&=\max\left\{1, \left\lfloor\frac{\sqrt{1+8\kappa_+/\Lambda}-1}{2}\right\rfloor\right\}. \label{eq:explicit-threshold}
\end{align}
Two smooth Einstein germs as in Theorem~\ref{thm:jets} with the same transverse packages through order $J_\gamma$ agree on a uniform two-sided whole-section collar in their common marked Gaussian-null normalization.
\end{theorem}
\begin{proof}
For every integer $n>J_\gamma$ and every trace-free $k$,
\[
\langle\mathcal L_nk,k\rangle_{L^2} \ge[n(n+1)\Lambda-2\kappa_\gamma]\|k\|_{L^2}^2>0.
\]
Lemma~\ref{nr:difference} therefore propagates the given finite matching to every order. Theorem~\ref{thm:jets} turns that full jet equality into smooth equality. Both steps are needed.
\end{proof}
Here matching through weighted order $J$ means matching $(\mu^{(n)},\beta^{(n+1)},\phi^{(n+2)})$ for $1\le n\le J$, in addition to the near-horizon data. Thus $J=1$ fixes the package $(\mu^{(1)},\beta^{(2)},\phi^{(3)})$; it is not ordinary metric first-jet matching.

This number is computable from the curvature endomorphism, with no unnamed energy constant. A certified upper bound for $\kappa_\gamma$ may be used in its place. In a class with $\kappa_\gamma/\Lambda\le B$, the same displayed integer with $\kappa_+/\Lambda$ replaced by $\max(0,B)$ works uniformly, without a uniform collar radius being asserted. The proof uses arbitrarily high smooth derivatives after the initial matching. Consequently \eqref{eq:explicit-threshold} is not a finite-$C^K$ regularity theorem.

The pointwise bound above can overestimate how many packages are needed. Define the finite resonance set and its terminal index by
\begin{equation}
\mathcal R_\gamma= \{n\in\mathbb Z:n\geq2,\ [2-n(n+1)]\Lambda \in\spec(\Delta_L|_{\operatorname{Sym}^2_0})\}, \qquad J_{\mathrm{spec}}=\max(\{1\}\cup\mathcal R_\gamma). \label{nr:spectral-threshold}
\end{equation}
Finiteness follows because the specified eigenvalues tend to $-\infty$, whereas the elliptic operator is bounded below.

\begin{corollary}[Matching through the last possible resonance]
\label{nr:finite-spectral}
The conclusion of Theorem~\ref{thm:finite-explicit} holds with $J_\gamma$ replaced by $J_{\mathrm{spec}}$. In particular, absence of every higher resonance makes the first transverse package sufficient. No assumption that the first kernel is parallel is needed for this comparison statement.
\end{corollary}
\begin{proof}
For each integer $n>J_{\mathrm{spec}}$ the homogeneous difference operator $\mathcal L_n$ has zero kernel. The induction in Theorem~\ref{thm:finite-explicit} therefore applies without using pointwise coercivity.
\end{proof}

If only a certified bound $\Delta_L|_{\operatorname{Sym}^2_0}\geq\ell\,\mathrm{Id}$ is available, the directly computable sufficient index
\begin{equation}
J(\ell)=\max\left\{1, \left\lfloor\frac{\sqrt{1+4\max(0,2-\ell/\Lambda)}-1}{2} \right\rfloor\right\} \label{nr:lower-spectral-threshold}
\end{equation}
works, since the shifted lower bound is strictly positive for every $n>J(\ell)$. The floor is essential when a lower bound vanishes at an integer indicial order: that order must be matched. The original curvature estimate is recovered by taking $\ell=2\Lambda-2\kappa_\gamma$. In particular a certified $\ell>-4\Lambda$ gives $J(\ell)=1$. The exact resonance set is the smallest sufficient terminal index provided by this homogeneous recursion. It is not claimed to be the minimal geometric data count: a resonant linear coefficient might fail a nonlinear compatibility condition and have no actual realization.

\label{sec:nonround}
We now identify the sections for which these finite-jet statements become an exhaustive classification. A potential first-order modulus is an element of $\ker\Delta_L$; a potential higher-order modulus requires a resonance of the shifted operator. The distinction is important: a spectral gap by itself says nothing about whether the first-order kernel is geometric. Related deformation formalisms have been developed in \cite{GSW,FontanellaGutowski2017}.

Let $\mathcal P_\gamma$ denote the space of global parallel symmetric 2-tensors on a compact connected positive Einstein section $(S,\gamma)$, with $\Ric(\gamma)=\Lambda\gamma$. Every tensor in $\mathcal P_\gamma$ belongs to $\ker\Delta_L$: its parallel self-adjoint endomorphism has parallel eigenspaces; the resulting local product has no mixed curvature, so $\mathring R k=\Lambda k$ and $\nabla k=0$. No sign assumption on sectional curvature is used here.

We shall use the following two spectral conditions:
\begin{align}
\ker\Delta_L&=\mathcal P_\gamma,\label{nr:kernel-condition}\\
[2-n(n+1)]\Lambda&\notin \spec(\Delta_L|_{\operatorname{Sym}^2_0}),\qquad n=2,3,\ldots. \label{nr:noresonance}
\end{align}
The restrictions act on all trace-free symmetric tensors, not only on transverse-traceless tensors. They are explicit hypotheses on the fixed section; positivity of its Ricci tensor alone does not imply them. Since $\Delta_L$ is a self-adjoint elliptic operator on a compact manifold, its spectrum is bounded below. Thus only finitely many of the numbers in \eqref{nr:noresonance} can meet the spectrum. In particular,
\begin{equation}
\inf\spec(\Delta_L|_{\operatorname{Sym}^2_0})>-4\Lambda \label{nr:convenient-gap}
\end{equation}
implies \eqref{nr:noresonance}, but must still be combined with \eqref{nr:kernel-condition}.

\begin{proposition}[Spectral formal classification]
\label{nr:formal}
Assume \eqref{nr:kernel-condition}--\eqref{nr:noresonance}. Every constant-trace first transverse coefficient of an Einstein germ with unwarped data $(\Lambda,0,\gamma)$ is parallel. Once it is fixed, all subsequent transverse packages are unique. If $\mathcal P_\gamma=\mathbb R\gamma$, these packages equal those of generalized extremal Kottler--de Sitter, including its Nariai limit.
\end{proposition}
\begin{proof}
Lemma~\ref{nr:first} places $\mu^{(1)}$ in $\ker\Delta_L$, so \eqref{nr:kernel-condition} makes it parallel. The companion coefficients are then $\beta^{(2)}=0$ and $\phi^{(3)}=-\Lambda\tr_\gamma\mu^{(1)}$. If the packages of two actual germs agree through order $n-1$, Lemma~\ref{nr:difference} makes their next difference an element of $\ker\mathcal L_n$ on trace-free tensors. Condition \eqref{nr:noresonance} sets this difference, and hence its entire package, to zero. Induction proves formal uniqueness. Actual realizations of all the permitted first data are constructed below.
\end{proof}

\begin{lemma}[Nonnegative sectional curvature supplies the hypotheses]
\label{nr:bochner}
If $\operatorname{Sect}(\gamma)\geq0$, then $\Delta_L\geq0$ and $\ker\Delta_L=\mathcal P_\gamma$. Consequently \eqref{nr:kernel-condition}--\eqref{nr:noresonance} hold.
\end{lemma}
\begin{proof}
Diagonalize a symmetric tensor $k$ at a point with eigenvalues $\lambda_i$. Since $\sum_jK_{ij}=\Lambda$,
\[
\Lambda|k|^2-\langle\mathring Rk,k\rangle =\tfrac12\sum_{i,j}K_{ij}(\lambda_i-\lambda_j)^2.
\]
Integration gives
\[
\langle\Delta_Lk,k\rangle_{L^2} =\int_S\left(|\nabla k|^2+ \sum_{i,j}K_{ij}(\lambda_i-\lambda_j)^2\right)\geq0.
\]
Equality implies $\nabla k=0$; the converse was proved above. For $n\geq2$, the added constant $[n(n+1)-2]\Lambda$ is strictly positive.
\end{proof}

\subsection{Exact realizations and smooth classification}

The recursion determines formal coefficients, but a classification requires actual solutions. Multiply warped Einstein metrics and their curvature equations are classical \cite{DobarroUnal2005}, as are singular initial-value problems for cohomogeneity-one Einstein metrics \cite{EschenburgWang2000}. We solve the realization problem by an analytic integral equation.

Let $P$ be a parallel self-adjoint endomorphism of $(TS,\gamma)$. Its eigenvalues $c_i$ are constant and its mutually orthogonal eigenbundles are parallel. Let $\gamma_i$ denote the associated parallel semidefinite metrics and $m_i$ their ranks, so $\gamma=\sum_i\gamma_i$. Locally this is a Riemannian product; the factor Ricci tensors are $\Lambda\gamma_i$. These objects are global on $S$, even if a global product decomposition requires a cover.

\begin{theorem}[Analytic multiwarp realization]
\label{nr:realization}
For every such parallel $P$ there is a unique analytic solution near $r=0$ within the ansatz
\begin{equation}
g_P=2\,dv\,dr+f(r)\,dv^2+\sum_i a_i(r)^2\gamma_i, \qquad f(r)=r^2F(r), \label{nr:ansatz}
\end{equation}
with $a_i(0)=1$, $a_i'(0)=c_i$, and $F(0)=\Lambda$. It satisfies the full equation $\operatorname{Ric}(g_P)=\Lambda g_P$ on a two-sided neighborhood of the whole compact section. Its first transverse coefficient is $2\gamma(P\cdot,\cdot)$. Here ``analytic'' refers to the $r$-dependence; the section tensors are the given smooth parallel tensors.
\end{theorem}
\begin{proof}
Put $H_i=a_i'/a_i$ and $H=\sum_i m_iH_i$. The base metric is $2\,dv\,dr+f\,dv^2$, with determinant $-1$ and inverse radial coefficient $-f$. Thus $\Box a_i=-(fa_i')'$ and $\langle da_i,da_j\rangle=-fa_i'a_j'$. The warped-product Ricci identities consequently give
\begin{align}
R_{rr}&=-H'-\sum_i m_iH_i^2,\label{nr:rr}\\
R_{vr}&=\tfrac12(f''+Hf'),\qquad R_{vv}=fR_{vr},\label{nr:vr}\\
R|_i&=\left[\Lambda a_i^{-2}+fH_i'+(f'+fH)H_i\right]a_i^2\gamma_i. \label{nr:internal}
\end{align}
The other mixed components vanish. It suffices initially to solve
\begin{equation}
f''+Hf'=2\Lambda,\qquad fH_i'+(f'+fH)H_i=\Lambda(1-a_i^{-2}). \label{nr:evolution}
\end{equation}

Let $D=r\partial_r$, $u_i=\log a_i$, $w_i=Du_i$, $z=F/\Lambda-1$, $v=Dz$, and $W=\sum_i m_iw_i$. Then \eqref{nr:evolution} is precisely the analytic autonomous system
\begin{align}
Du_i&=w_i,\nonumber\\
Dw_i&=\frac{1-e^{-2u_i}}{1+z} -\left(1+\frac{v}{1+z}+W\right)w_i,\nonumber\\
Dz&=v,\label{nr:BB}\\
Dv&=-2z-3v-2W(1+z)-Wv.\nonumber
\end{align}
Write its right side as $AY+N(Y)$, where $Y=(u_i,w_i,z,v)$ and $N(Y)=O(|Y|^2)$ is analytic at zero. The spectrum of $A$ consists of $1,-2$ for each factor and the two additional values $-1,-2$. The $+1$ eigenspace is semisimple and has dimension equal to the number of factors. For the chosen $c_i$, its associated vector is
\[
e=(c_i,c_i,-C_0/3,-C_0/3),\qquad C_0=\sum_i m_ic_i.
\]
In particular $Ae=e$.

Set $Y=r(e+Z)$. Then
\[
DZ=BZ+rQ(r,Z),\qquad B=A-I,
\]
where $Q$ is analytic near $(0,0)$. The spectrum of $B$ is $0,-2,-3$, and zero is semisimple. Consequently $t^{-B}$ is bounded on $0<t\leq1$; any Jordan blocks at negative eigenvalues merely add bounded terms of the form $t^j(\log t)^q$ with $j>0$. On a sufficiently small complex disk, define
\begin{equation}
(\mathcal TZ)(r)=r\int_0^1 t^{-B} Q(rt,Z(rt))\,dt. \label{nr:contraction}
\end{equation}
Use the Banach space of functions holomorphic in the disk, continuous on its closure, with $Z(0)=0$, equipped with the supremum norm. On a fixed small norm ball, analyticity bounds both $Q$ and its $Z$-derivative. If the disk radius is $R$, the image norm is at most $R C_B\sup|Q|$ and the Lipschitz constant is at most $R C_B\sup|D_ZQ|$, where $C_B=\int_0^1\|t^{-B}\|dt<\infty$. Taking $R$ small makes $\mathcal T$ preserve the ball and be a contraction. Its fixed point is holomorphic and solves the differential system by differentiating \eqref{nr:contraction}. Real coefficients and uniqueness make it real on real $r$. Conversely any analytic solution with the specified first coefficient has the same integral formula, since the homogeneous endpoint term vanishes. This proves analytic existence and uniqueness; no unsupported summation of a formal series has been used.

It remains to verify the apparently missing $rr$ constraint. Define $Q_0=H'+\sum_i m_iH_i^2$ and write $E_i=fH_i'+(f'+fH)H_i-\Lambda(1-a_i^{-2})$ for the residual of the $i$-th internal equation. Direct differentiation gives
\[
\left(\sum_i m_iE_i\right)'+2\sum_i m_iH_iE_i=fQ_0'+2(f'+fH)Q_0+H(f''+Hf'-2\Lambda).
\]
Since all $E_i$ vanish and $f''+Hf'=2\Lambda$ by \eqref{nr:evolution}, this yields
\[
fQ_0'+2(f'+fH)Q_0=0.
\]
Equivalently,
\[
\left(f^2\prod_i a_i^{2m_i}Q_0\right)'=0.
\]
The constructed solution has bounded $Q_0$ and $f=\Lambda r^2+O(r^3)$, so this constant is zero on each side. Hence $Q_0=0$ there and by continuity also at $r=0$. Equations \eqref{nr:rr}--\eqref{nr:internal} now prove the full Einstein equation. For small real $|r|$, all $a_i$ are positive and $F>0$. The $(v,r)$ block has determinant $-1$, so the metric is a smooth Lorentzian metric through the horizon; compactness of $S$ supplies a uniform collar. The coefficients are common analytic functions on both sides, not two independently glued solutions.
\end{proof}

\begin{theorem}[Smooth classification from the full tensor spectrum]
\label{nr:smooth}
Let $d=m+2\geq4$ and $\Lambda>0$. Fix a smooth compact connected Einstein section $(S^m,\gamma)$ without boundary, with $\Ric(\gamma)=\Lambda\gamma$, satisfying \eqref{nr:kernel-condition}--\eqref{nr:noresonance}. Let a smooth $d$-dimensional solution of $\Ric(g)=\Lambda g$ have an actual marked section identified with $S$, a nonvanishing normalized degenerate Killing generator, and unwarped horizon data $(\Lambda,0,\gamma)$. Its germ is, after the constant-trace cut normalization, smoothly locally isometric on a whole two-sided collar to exactly the multiwarp germ supplied by its parallel first transverse coefficient. In particular, if $\mathcal P_\gamma=\mathbb R\gamma$, the only such germs are generalized Kottler--de Sitter or generalized Nariai. The isometry preserves the normalized generator and normalized marking; the original marked cut is retained as its graph. Nonnegative sectional curvature is one sufficient condition by Lemma~\ref{nr:bochner}, and is not required by this theorem.
\end{theorem}
\begin{proof}
Proposition~\ref{nr:formal} shows that the normalized first coefficient is parallel. Theorem~\ref{nr:realization} supplies an actual comparison solution with the same first package: its shift vanishes, and both scalar cubic coefficients are $-\Lambda\operatorname{tr}\mu^{(1)}$. Proposition~\ref{nr:formal} forces equality of every higher package. Theorem~\ref{thm:jets} then yields the stated isometry. The final invocation is essential: analytic realization alone is not uniqueness among smooth germs.
\end{proof}

\subsection{Spectral certificates and examples}

The spectral conditions can be certified without a pointwise sectional-curvature sign. The next examples lie beyond the preceding Bochner argument.

\begin{lemma}[Positive K\"ahler--Einstein spectral certificate]
\label{nr:kahler}
Let $(S,\gamma,J)$ be compact K\"ahler--Einstein with $\Ric(\gamma)=\Lambda\gamma$, $\Lambda>0$. Then $\Delta_L\geq0$ on the full symmetric-tensor bundle, and its kernel identifies with the real harmonic $(1,1)$-forms by $k\mapsto k(J\cdot,\cdot)$. Consequently $J_{\mathrm{spec}}=1$, without a topological restriction. If $b_2(S)=1$, then $\ker\Delta_L=\mathbb R\gamma$, so both \eqref{nr:kernel-condition} and \eqref{nr:noresonance} hold.
\end{lemma}
\begin{proof}
Write $k=k_H+k_A$ for its Hermitian and anti-Hermitian parts. In the normalization \eqref{nr:indicial}, the canonical spin$^c$ Weitzenb\"ock identity of Dai--Wang--Wei \cite[Theorem~2.6 and the identity immediately following its proof]{DWW2007} reads
\begin{equation}
\langle\Delta_Lk,k\rangle_{L^2} =\|\mathcal D\Phi(k)\|_{L^2}^2+2\Lambda\|k_A\|_{L^2}^2. \label{nr:kahler-energy}
\end{equation}
Here $\Phi$ is the parallel isometric map from symmetric tensors to spinor-valued one-forms induced by the canonical parallel spin$^c$ spinor, and $\mathcal D$ is the twisted Dirac operator. The cited identity holds for arbitrary symmetric tensors; no divergence or trace condition is imposed. It proves nonnegativity and that a kernel tensor is Hermitian. For a Hermitian tensor, $\alpha(X,Y)=k(JX,Y)$ is a real $(1,1)$-form. The parallel bundle identification intertwines $\Delta_L$ with the Hodge Laplacian:
\[
(\Delta_Lk)(J\cdot,\cdot)=\Delta_d\alpha.
\]
This follows by comparing their rough-Laplacian and curvature terms and using $\nabla J=0$; it fixes in particular the absence of a $2\Lambda$ shift on the Hermitian summand. Hence the kernel is exactly the harmonic real $(1,1)$-forms. If $b_2=1$, the nonzero parallel K\"ahler form spans this harmonic space, and its tensor is $\gamma$. The remaining statements follow from the positive shift in $\mathcal L_n$, $n\geq2$.
\end{proof}

\begin{corollary}[A homogeneous section beyond nonnegative curvature]
\label{nr:g2}
Let $Z=G_2/U(2)_+$ be the complex contact fivefold which is the twistor space of $G_2/SO(4)$, equipped with its invariant K\"ahler--Einstein metric scaled to $\Ric=\Lambda\gamma$. Every smooth twelve-dimensional Einstein germ with unwarped horizon data $(\Lambda,0,\gamma)$ is a generalized extremal Kottler--de Sitter or generalized Nariai germ after constant-trace cut normalization, on a whole two-sided collar. It is determined by its first transverse package. The metric $\gamma$ has a negative sectional direction.
\end{corollary}
\begin{proof}
The positive invariant K\"ahler--Einstein metric, simple connectedness, $b_2(Z)=1$ and $c_1(Z)=3L$ for a primitive positive integral class $L$ are recorded in \cite[Sections~1--2]{KotschickThung2020}. Lemma~\ref{nr:kahler} and Corollary~\ref{nr:smooth} give the classification. Suppose that all sectional curvatures were nonnegative. The K\"ahler curvature identity then gives nonnegative holomorphic bisectional curvature. Mok's theorem \cite[Theorem~1]{Mok1988}, together with $b_2=1$ and positive Ricci curvature, makes $Z$ complex projective space or an irreducible compact Hermitian symmetric space. In complex dimension five the latter list consists of the quadric $Q^5$; the other possibility is $\mathbb{CP}^5$. Their first Chern classes have divisibility five and six, respectively, contradicting $c_1(Z)=3L$. This proves the negative-sectional-curvature assertion.
\end{proof}

The subgroup notation matters: the other nonconjugate quotient $G_2/U(2)_-$ is $Q^5$, and would not give this strict extension. The contact fivefold $Z$ has two irreducible isotropy summands, whereas the quadric presentation has three \cite[Section~1]{KotschickThung2020}. The following independent example gives the same strictness in ambient dimension eight.

\begin{corollary}[Cubic-threefold horizons]
\label{nr:cubic}
Let $X\subset\mathbb{CP}^4$ be any smooth cubic hypersurface and let $\gamma$ be a positive K\"ahler--Einstein metric on $X$, rescaled so that $\Ric(\gamma)=\Lambda\gamma$. Every smooth eight-dimensional vacuum Einstein germ with unwarped horizon data $(\Lambda,0,\gamma)$ is, after constant-trace cut normalization, a generalized extremal Kottler--de Sitter or generalized Nariai germ on a whole two-sided collar. Its first transverse package determines it. Moreover $\gamma$ has a negative sectional curvature at some point and plane, so this classification is not covered by $\operatorname{Sect}\geq0$.
\end{corollary}
\begin{proof}
The required K\"ahler--Einstein metric exists by Liu--Xu \cite[Corollary~1.2(1)]{LiuXu2019}. In particular one may take the explicit smooth variety $z_0^3+z_1^3+z_2^3+z_3^3+z_4^3=0$; no explicit coordinate formula for its Einstein metric is needed. The Lefschetz hyperplane theorem gives $\pi_1(X)=0$, $b_1=0$ and $b_2=1$. Lemma~\ref{nr:kahler} and Corollary~\ref{nr:smooth} therefore give the ambient classification.

To establish the curvature assertion, let $H$ be the restricted hyperplane class. The tangent-normal exact sequence gives
\[
c(TX)=\frac{(1+H)^5}{1+3H} =1+2H+4H^2-2H^3, \qquad \int_XH^3=3.
\]
Thus $\chi(X)=-6$. Poincar\'e duality and the preceding Lefschetz calculation give $\chi(X)=4-b_3(X)$, and hence $b_3(X)=10$. If $\gamma$ had nonnegative sectional curvature everywhere, the K\"ahler curvature identity would give nonnegative holomorphic bisectional curvature: for real tangent vectors the bisectional curvature numerator is the sum of the sectional-curvature numerators of the planes $(X,Y)$ and $(X,JY)$. Mok's uniformization theorem \cite[Main Theorem]{Mok1988}, applied to the compact simply connected K\"ahler manifold $X$, then makes $X$ biholomorphic to a product of complex projective spaces and irreducible compact Hermitian symmetric spaces. Those factors have Schubert cell decompositions in even real dimensions, so their product has vanishing odd Betti numbers. This contradicts $b_3(X)=10$. Thus a negative sectional direction must occur.
\end{proof}

The spectral certificate uses established K\"ahler identities and existence theorems; the new conclusion is the smooth ambient exhaustion for these horizon sections. It neither constructs a new K\"ahler--Einstein metric nor asserts that positive Einstein metrics in general satisfy \eqref{nr:kernel-condition}. For example, on a positive K\"ahler--Einstein section with $b_2>1$, additional harmonic $(1,1)$-forms can yield nonparallel first-jet kernel tensors. Their nonlinear realization is a different problem.

For a single factor of dimension $m\geq2$, the $rr$ equation gives $a''=0$, hence $a=1+cr$. Solving the remaining equation yields
\begin{equation}
f_c(r)=\frac{\Lambda}{c^2(m+1)} \left[(1+cr)^2+\frac{2}{m-1}(1+cr)^{1-m} -\frac{m+1}{m-1}\right],\quad c\ne0. \label{nr:kottler}
\end{equation}
Its continuous $c=0$ limit is $\Lambda r^2$. This is the standard generalized extremal Kottler family in a horizon normalization; Einstein-horizon black-hole and cosmological metrics have long been studied \cite{GibbonsHartnollPope2003}. The first two section jets are $2c\gamma$ and $2c^2\gamma$.

On $\mathbb{CP}^2$, normalize Fubini--Study to $\operatorname{Ric}=\Lambda\gamma$. If its holomorphic sectional curvature is $4c_0$, then $c_0=\Lambda/6$ and
\[
\mathring R k=c_0\big[(\operatorname{tr}k)\gamma-k+3JkJ^T\big].
\]
The trace-free Hermitian and anti-Hermitian summands have curvature action eigenvalues $\Lambda/3$ and $-2\Lambda/3$. Therefore $\Delta_L\geq4\Lambda/3$ on trace-free symmetric tensors. There is no first or higher trace-free transverse resonance, and Corollary~\ref{nr:smooth} gives the six-dimensional generalized Kottler/Nariai classification. For comparison, the scalar spectrum is $\frac{2\Lambda}{3}k(k+2)$, $k=0,1,\ldots$; its first positive value is $2\Lambda$. The first Hodge one-form value is also $2\Lambda$, realized by gradients from the first scalar eigenspace and by Killing one-forms. The latter lower bound follows on coexact forms from the Killing identity, and on exact forms from the scalar spectrum. These spectral facts are checks; the classification uses the explicit full-tensor curvature bound.

On $S^2\times S^2$ with each factor normalized to Einstein constant $\Lambda$, the curvature action on trace-free symmetric tensors has values $\Lambda,-\Lambda,0$ on, respectively, relative block scaling, within-block trace-free tensors, and mixed tensors. The first-jet space is $\mu^{(1)}=2c_1\gamma_1+2c_2\gamma_2$, and every pair $(c_1,c_2)$ is realized by Theorem~\ref{nr:realization}. The relative-scaling tensor, familiar from Einstein-metric stability \cite{Schwahn2022}, $\gamma_1-\gamma_2$ satisfies $\Delta_L(\gamma_1-\gamma_2)=0$, so it is a genuine order-one resonance. In particular $c_1=-c_2\ne0$ gives a convergent nonlinear pure trace-free departure from the product horizon. No higher-order obstruction removes it. For $c_1=c$, $c_2=-c$, put $T=\gamma_1-\gamma_2$. The first few coefficients of the convergent solution are
\begin{align}
\mu(r)&=\gamma+2crT+c^2r^2\gamma +\tfrac4{15}c^3r^3T+O(r^4),\nonumber\\
F(r)&=\Lambda\left(1+\tfrac23c^2r^2+O(r^4)\right). \label{nr:TT-jets}
\end{align}
Thus $\mu^{(2)}=2c^2\gamma$, $\mu^{(3)}=\frac85c^3T$, $\phi^{(3)}=0$ and $\phi^{(4)}=16\Lambda c^2$. These are direct checks of the first nonlinear constraints, not the convergence argument, which is \eqref{nr:contraction}. The subfamily $c_2=0$, $c_1\ne0$ is explicitly the product of four-dimensional extremal Schwarzschild--de Sitter with a fixed Einstein $S^2$. This supplies an actual counterexample to a Kottler-only ambient classification on the product section; the known direct-product metric itself is not a new solution.

The distinction from a nonconstant singly warped Kottler metric is also invariant under unmarked local isometries.  Write that metric as $B+a^2\gamma$, where $B=2\,dv\,dr+f\,dv^2$ and $a=1+cr$, $c\ne0$. A direct calculation gives $\nabla_B^2a=-\tfrac12cf'B$. At a nearby point with $r\ne0$, one has $f'\ne0$; the mixed curvature endomorphisms are therefore nonzero scalar multiples of every elementary base--fiber skewendomorphism $U\wedge X$, $U\in TB$, $X\in TS$. Their Lie brackets generate the full Lorentz algebra, so the restricted holonomy is full and there is no parallel spectator factor. In contrast, extremal $\mathrm{SdS}_4\times S^2$ has a parallel rank-two spacelike factor.  The generalized Nariai limit on $S^2\times S^2$ is locally symmetric, whereas $\mathrm{SdS}_4\times S^2$ has the nonconstant curvature invariant contributed by the four-dimensional Weyl term $48M^2/R^6$. Here $M\ne0$ is the four-dimensional Schwarzschild mass and $R$ its area radius.  It is consequently inequivalent to that limit as well.

The scalar spectrum on this product is $\Lambda[\ell_1(\ell_1+1)+\ell_2(\ell_2+1)]$. Positive Ricci curvature eliminates $H^1$, but does not eliminate the parallel trace-free tensor. This distinction explains why a cohomology argument cannot remove the extra geometric modulus. The parameters classify normalized, marked germs. More precisely, if $K'=aK$ with constant $a\ne0$, the coordinates $v'=v/a$, $r'=ar$ preserve $2\,dv\,dr$ and change $c_i$ to $c_i/a$. Restrict $a>0$ if the generator's time orientation is part of the marking. Section isometries conjugate the parallel endomorphism and can permute isometric factors. A cut change acts on the first tensor by a Hessian in the convention of Lemma~\ref{nr:first}. If both cuts have constant trace, integrating the trace transformation shows that this Hessian has zero Laplacian; compact connectedness then makes its generating function constant. Thus no nonconstant residual cut shift changes the normalized slopes. On a finite quotient of a product section, only slope assignments invariant under the deck action descend. These are exactly the global parallel endomorphisms already used in the realization theorem. The holonomy argument above establishes genuine unmarked distinctness from the singly warped family for the spectator example; it is not a claim that all ratios of nonzero slopes classify completely unmarked Lorentzian germs. Such a moduli assertion would require control of isometries that need not preserve the chosen horizon or its Killing generator.

For both examples $J_\gamma=1$ in Theorem~\ref{thm:finite-explicit}. On $S^2\times S^2$ this first jet is necessary: the intrinsic horizon metric alone does not fix the two slopes. For a round section, Corollary~\ref{nr:smooth} recovers the following smooth version of the Katona--Lucietti classification.
\begin{corollary}[Round-section rigidity]\label{pos:theorem}
In every $d\ge4$, a smooth Einstein germ with $\Lambda>0$ and an unwarped compact maximally symmetric horizon section is locally extremal Schwarzschild--de Sitter or Nariai. The comparison is smooth on both sides of a uniform neighborhood of the whole section. The original cut is retained as its graph after constant-trace normalization.
\end{corollary}

\subsection{Finite metric regularity}\label{posfinite:section}

The explicit index $J_\gamma$ in Theorem~\ref{thm:finite-explicit} counts the transverse packages needed to determine a \emph{smooth} germ. It is not a differentiability threshold: its proof propagates finite matching through arbitrarily high orders. For a metric with finitely many derivatives, the available Taylor decay must instead exceed the growth rates in the terminal energy estimates. We give such an argument first for round sections. The resulting threshold depends only on dimension, though its value is implicit in fixed cylinder energy constants. In the smooth setting, the comparison below is weaker than Theorem~\ref{thm:finite-explicit}; its purpose is to control finite regularity.

\begin{proposition}[Finite-decay comparison for round sections]
\label{posfinite:jets}
For every $d\geq4$ there is an integer $n_0(d)$ with the following property. Let $g$ be a smooth Einstein germ as in Corollary~\ref{pos:theorem}, and let $g_0$ be a Schwarzschild--de Sitter or Nariai model in the same normalized Gaussian-null gauge. If the transverse derivatives of $(F,h,\mu)$ agree through order $n_0(d)$, the two germs are locally isometric on a neighborhood of the whole compact section. The matching order is independent of $\Lambda$, the first transverse constant $C$, and the spherical-space-form quotient.
\end{proposition}

\begin{proof}
Rescale so that $\Lambda=1$, and lift the section to its round universal cover $(S^m,\gamma_*)$, where $m=d-2$ and $\Ric(\gamma_*)=\gamma_*$. The limiting quotient fields are
\[
q_*=-\dd t^2+\gamma_*,\qquad u_*=-t,\qquad W_*=\partial_t.
\]
Set $s=\lfloor m/2\rfloor+6$. Fix a small coefficient neighborhood of these fields, controlling $q,q_0$ in $C^{s+4}$ and $\dd u,\dd u_0$ in $C^{s+3}$ relative to $b=\dd t^2+\gamma_*$. Choose it small enough to give fixed lapse, inverse-metric and spacelike-slice bounds. Also fix a small displacement-energy ball for the exponential-map gauge.

The energy proof of Lemma~\ref{pos:terminal-gauge}, on this fixed round cylinder and fixed coefficient class, has a constant $c_G=c_G(d,s)$:
\[
|\partial_t e_s|\leq c_Ge_s+C\|f(t)\|_{H^s}.
\]
All constants here depend only on the prescribed finite coefficient bounds, the fixed round Sobolev estimates, and the fixed displacement range. Choose likewise constants $c_M(d)$ and $c_E(d)$ for the square-root Maxwell energy and the basic reduced metric/scalar comparison energy. The latter requires only a fixed low-order coefficient class: $C^2$ bounds on the pushed metric and positive-order lapse derivatives, together with $C^3$ background bounds, suffice; no $C^{s+4}$ bound on the pushed fields is required. Let $n=n_0(d)$ be an integer with
\begin{equation}
n>\max\{c_G(d,s),c_M(d),c_E(d),s+5,3\}. \label{posfinite:threshold}
\end{equation}

Fixing the coefficient classes before choosing the germ makes these constants dimension-only. Every germ enters the fixed input class on a sufficiently late tail, because its quotient fields converge with the required derivatives to $(q_*,\dd u_*)$. The parameter $C$ changes the starting time, not the limit. Working on the round cover avoids quotient injectivity-radius or Sobolev-constant dependence. Finite-terminal gauge solutions are deck-equivariant by the natural equation, identity terminal data and finite-time Cauchy uniqueness; their limit descends.

Taylor's theorem applied to the assumed matching gives, with the finite number of derivatives used here,
\begin{equation}
q-q_0,\ u-u_0=O(e^{-(n+1)t}),\qquad A-A_0,\ \Omega=O(e^{-nt}). \label{posfinite:rates}
\end{equation}
The loss of one power in the connection is essential: $\phi-\phi_0=O(r^{n+3})$, $\beta=O(r^{n+2})$, whereas $A_t-A_{0,t}=O(r^n)$ and $A_a=O(r^n)$.

The Maxwell comparison gives $e_M(t)\leq e^{c_M(T-t)}Be^{-nT}$, and hence $\Omega=0$. For the terminal gauge, the identity defect satisfies $\|f(t)\|_{H^s}\leq Be^{-(n+1)t}$, so
\begin{equation}
e_s(t)\leq \frac{B'}{n+1-c_G}e^{-(n+1)t} \label{posfinite:gauge-estimate}
\end{equation}
uniformly for all finite terminal times. A sufficiently late starting time closes the fixed bootstrap. The choice of $s$, the equation solved for $\partial_t^2Z$, and compact Sobolev embeddings give subsequential convergence with at least three classical derivatives on compact slabs. More explicitly, $Z$ and its first four time derivatives are controlled, with the same decay rate, in $H^{s+1},H^s,H^{s-1},H^{s-2},H^{s-3}$, respectively. The last three bounds follow by solving the equation for $Z_{tt}$ and differentiating twice; the mixed Taylor bounds control $f_t,f_{tt}$ and the required coefficient derivatives. The limit is a gauge diffeomorphism with the corresponding finite derivative decay. In particular its pushed fields converge in $C^2$ to the limiting fields, and enter the already-fixed low-order comparison class on a still later tail.

The exact diagonal reduced system then yields $e_E(t)\leq e^{c_E(T-t)}B''e^{-(n+1)T}$, so the pushed fields equal the model. The closed connection difference has its global primitive $f=-\int_t^\infty (A-Y^*A_0)_t\,\dd t$, now of order $r^n$. The lifted map consequently satisfies
\[
Y^t-t=O(r^{n+1}),\quad Y^S-x=O(r^{n+1}),\quad r'-r=O(r^{n+2}),\quad v'-v=O(r^n).
\]
Since $n>3$, the two side maps glue as a $C^2$ isometry with identity differential on the normalized horizon. Geodesic uniqueness in common Gaussian-null gauge then makes this map the identity.
\end{proof}

\begin{corollary}[Finite metric regularity]
\label{posfinite:regularity}
In Corollary~\ref{pos:theorem}, retain a smooth horizon, Killing generator and marked section, but assume only $g\in C^{K(d)}$, with the Einstein equation interpreted classically. A sufficient dimension-only threshold is
\begin{equation}
K(d)=n_0(d)+\left\lfloor\frac{d-2}{2}\right\rfloor+26. \label{posfinite:K}
\end{equation}
The same whole-section local classification holds, with a $C^{K(d)+1}$ local isometry in the original smooth atlases.
\end{corollary}

\begin{proof}
Writing $s=\lfloor(d-2)/2\rfloor+6$, the displayed budget is $K=n_0+s+20$. We give conservative coordinate losses explicitly. A $C^K$ metric gives a $C^{K-1}$ geodesic flow and therefore $C^{K-2}$ coefficients in initial Gaussian-null coordinates. Round developing coordinates for the induced constant-curvature metric can be constructed using its exponential map, leaving at least $C^{K-4}$ spacetime coefficients. The trace of the first transverse derivative is then $C^{K-5}$.

On the smooth round cover, the mean-zero Poisson equation defining the constant-trace cut has a solution of class at least $C^{K-4}$, by $W^{K-3,p}$ elliptic regularity and Sobolev embedding for large finite $p$. The new null initial vector involves one cut derivative; the new Gaussian-null map is at least $C^{K-5}$ and its metric coefficients are at least $C^{K-6}$.

The finite Katona--Lucietti recursion through order $n_0$ uses at most $n_0+2$ ordinary metric derivatives, including its tangential derivatives. It therefore supplies the needed finite jet. There is no accumulating elliptic loss in that finite recursion. The remaining regularity is more than sufficient for Taylor estimates through order $n_0+1$ with mixed derivatives through $s+3$: a conservative requirement is $n_0+s+6$ derivatives of the normalized metric, whereas $K-6=n_0+s+14$. This also covers division by $r^2$ and the finite coefficient bounds in the gauge construction.

The proof of Proposition~\ref{posfinite:jets} thus applies at finite regularity and gives a $C^2$ local isometry, including across the horizon. In original coordinates it obeys the connection transformation equation
\[
\partial_i\partial_j\Psi^\alpha =\Gamma(g)^k{}_{ij}\partial_k\Psi^\alpha -\Gamma(g_0)^\alpha{}_{\beta\gamma}(\Psi) \partial_i\Psi^\beta\partial_j\Psi^\gamma .
\]
Since $\Gamma(g)\in C^{K-1}$ and the target metric is smooth, successive differentiation bootstraps $\Psi$ from $C^2$ to $C^{K+1}$.
\end{proof}

The bound $K(d)$ is sufficient rather than sharp. Its energy constants are fixed before the germ is chosen; the resulting collar may still depend on the germ. Throughout this finite-regularity statement the horizon, generator and section remain smooth, while the metric is assumed only $C^{K(d)}$.

On a fixed nonround positive-Einstein section, the same approach gives a threshold depending on that section and its cylinder estimates.

\begin{corollary}[Finite regularity with a fixed smooth Einstein section]
\label{posfinite:fixed-gamma}
Fix a smooth compact positive-Einstein metric $\gamma$ and a smooth exact unwarped Einstein germ $g_0$ with this section. There is an integer $N_\gamma\ge J_\gamma$, depending on a fixed cylinder coefficient class and Sobolev constants for $\gamma$, such that the following holds. Let $g\in C^{K_\gamma}$ be a classical Einstein metric with smooth horizon, generator and section, and with the induced section metric identified with the fixed smooth $\gamma$. Put
\[
s=\lfloor m/2\rfloor+6,\qquad K_\gamma=N_\gamma+s+20.
\]
If its normalized transverse packages agree with $g_0$ through order $J_\gamma$, then the germs are isometric across both sides of the whole section, by a $C^{K_\gamma+1}$ map in the original smooth atlases. If $\gamma$ satisfies the spectral hypotheses \eqref{nr:kernel-condition}--\eqref{nr:noresonance}, every such finite-regularity unwarped germ is a member of the family in Theorem~\ref{nr:realization}, after the constant-trace cut normalization.
\end{corollary}
\begin{proof}
Replace the round limiting cylinder in the preceding proof by $-\Lambda^{-1}dt^2+\gamma$. Fix its coefficient and displacement neighborhoods first, and then their constants $c_G,c_M,c_E$. Choose
\[
N_\gamma\ge J_\gamma,\qquad N_\gamma>\max\{c_G,c_M,c_E,s+5,3\}.
\]
The individual solutions affect only the time at which they enter these fixed classes. Apply Lemma~\ref{nr:difference} and its coercivity only finitely many times, from $J_\gamma+1$ through $N_\gamma$. The matched packages then give the finite Taylor bounds \eqref{posfinite:rates}. The smooth reference is twist-free by Theorem~\ref{thm:frobenius}; the finite-decay Maxwell estimate in the preceding proof eliminates the twist of $g$. The same finite-terminal gauge, reduced comparison energy and primitive construction yield a $C^2$ isometry across the horizon. The derivative budget remains $K_\gamma-6=N_\gamma+s+14$, exceeding the mixed Taylor and coefficient requirements. The connection equation with smooth target $g_0$ gives the stated regularity in the original atlases.

Under the spectral hypotheses the finite first-order equations place the normalized first coefficient in $\ker\Delta_L=\mathcal P_\gamma$, so it is parallel. Choose its smooth realization $g_P$. The higher-order nonresonance in Proposition~\ref{nr:formal} matches the remaining finitely many packages through $N_\gamma$, after which the preceding finite-decay argument applies directly. The proof uses only finite recursion and finite-decay continuation at the stated differentiability.
\end{proof}

The two indices serve different purposes. The matching index $J_\gamma$ is explicit from the curvature endomorphism, whereas the finite-regularity index $K_\gamma$ also depends on the fixed section's Sobolev and cylinder energy constants. A uniform regularity threshold across a varying family of sections would require uniform control of those constants.

\section{Einstein--Maxwell Horizons}\label{sec:einstein-maxwell}
\label{em:section}

Electromagnetic fields change the continuation problem in a specific way: the twist equation acquires a source, while electric and magnetic fluxes need not admit globally defined potentials. Both features are visible in the first transverse equations studied in \cite{LiLucietti2019,Kolanowski2022,Katona2024}. We retain the three reduced one-forms and compare field strengths directly. This gives a coupled smooth theorem under exactly the same orbit-slicing condition as in vacuum.

\subsection{Coupled smooth continuation}

We use signature $(-+++)$ and the normalization
\begin{equation}
R_{\alpha\beta}(g)=\Lambda g_{\alpha\beta} +2\left(\mathcal F_{\alpha\rho}\mathcal F_\beta{}^\rho -\frac14g_{\alpha\beta}\mathcal F_{\rho\sigma}\mathcal F^{\rho\sigma}\right), \qquad d\mathcal F=0,\quad d\star_g\mathcal F=0. \label{em:equations}
\end{equation}
The Killing field is assumed to preserve $\mathcal F$.

\begin{theorem}[Coupled smooth jet determination]
\label{em:jet}
Let two smooth four-dimensional solutions of \eqref{em:equations}, with the same $\Lambda$ and nonvanishing Killing generators satisfying $\mathcal L_{K_i}\mathcal F_i=0$, have identified compact connected boundaryless spacelike sections of degenerate Killing horizons.  Fix the generator, the marked section, and a common Gaussian-null normalization.  Write
\begin{align}
g_i&=r^2F_i\,dv^2+2\,dv\,dr+2r h_{i,a}\,dv\,dx^a +\mu_{i,ab}\,dx^a dx^b,\label{em:gnc}\\
\mathcal F_i&=\Psi_i\,dv\wedge dr+r\,dv\wedge W_i +dr\wedge Z_i+B_{S,i}.\label{em:emgnc}
\end{align}
Assume all transverse jets of both tensors agree at the section.  Suppose there is a single global smooth function $\psi$ on the common section such that the horizon data satisfy
\begin{equation}
F_0>|h_0+d\psi|_{\mu_0}^{\,2} \quad\hbox{everywhere on the section}. \label{em:strict-sign}
\end{equation}
Then the two solutions are smoothly isometric, including their Maxwell fields, on a two-sided neighborhood of the whole section.  The isometry preserves the normalized generator and marking.  In the common normalized Gaussian-null coordinates the coefficient functions themselves agree on a smaller collar.
\end{theorem}

The analytic hypothesis is the geometric inequality \eqref{em:strict-sign}. In particular, $h_0=0$, $F_0=a>0$ implies it. For a compact electrovacuum horizon the $vr$ equation reads
\[
F_0=\Lambda+\tfrac12|h_0|^2-\tfrac12\operatorname{div}h_0 -\Psi_0^2-\tfrac12|B_{S,0}|^2,
\]
where $|B_{S,0}|^2=(B_{S,0})_{ab}(B_{S,0})^{ab}$ in this formula. The weight $e^{4\psi}$ is chosen so that integration by parts turns $-\frac12\operatorname{div}h_0$ into $2\langle h_0,d\psi\rangle$, cancelling the cross term in $|h_0+d\psi|^2$. Multiplication of the strict inequality by this weight therefore gives
\[
\Lambda\int_S e^{4\psi}>\int_S e^{4\psi} \bigl(\tfrac12|h_0|^2+|d\psi|^2+\Psi_0^2+\tfrac12|B_{S,0}|^2\bigr),
\]
so $\Lambda>0$ is again necessary. The theorem compares two actual solutions; it does not prescribe existence from arbitrary formal data. Orientations may be chosen on the collars; alternatively, the argument can be made on their orientation covers and descends by uniqueness.

On either punctured side put $r=\sigma e^{-t+\psi(x)}$, $\sigma\in\{-1,1\}$, and set
\begin{align}
V&=|r|\sqrt F,\quad u=\log V,\quad p=du,\nonumber\\
A&=\frac{dr+rh}{r^2F},\quad \vartheta=dv+A,\quad q=\mu-F^{-1}(-dt+d\psi+h)^2. \label{em:variables}
\end{align}
Thus $g=V^2\vartheta^2+q$.  Define horizontal tensors
\begin{equation}
E=V^{-1}\iota_K\mathcal F,\qquad B=\mathcal F-\vartheta\wedge\iota_K\mathcal F,\qquad C=V\,dA,\qquad H=\star_q B,\qquad J=\star_q C. \label{em:regular-fields}
\end{equation}
Here $q$ has Lorentzian signature $(-++)$, its volume orientation is fixed by $\operatorname{vol}_g=V\vartheta\wedge\operatorname{vol}_q$, and $\star_q^2=-1$ on one- and two-forms.  Hence $B=-\star_q H$, $C=-\star_q J$ and $\mathcal F=V\vartheta\wedge E-\star_q H$.

The form \eqref{em:emgnc} follows from the horizon hypotheses rather than being an extra alignment condition. A Killing horizon has zero expansion and shear, so Raychaudhuri's equation gives $\operatorname{Ric}(K,K)=0$ there. The Einstein--Maxwell equation makes this twice the squared spatial norm of $\iota_K\mathcal F$ on the section. Thus $\mathcal F_{va}|_{r=0}=0$, and smooth division supplies the one-form $W$ in \eqref{em:emgnc}. The coefficient $Z$ is unrestricted at this step.

These fields have bounded derivatives of every order in a fixed product metric on the cylinder.  This is verified directly, rather than assumed: with $\alpha=F^{-1}(-dt+d\psi+h)$,
\begin{align}
p&=-dt+d\psi+\tfrac12d\log F,& C&=\sigma\sqrt F\,((dt-d\psi)\wedge\alpha+d\alpha),\label{em:regular-explicit}\\
E&=\frac{\sigma}{\sqrt F}[-\Psi(dt-d\psi)+W],\nonumber\\
B&=B_S-r(dt-d\psi)\wedge Z\nonumber\\
&\quad+F^{-1}\{(dt-d\psi)\wedge(W-\Psi h)-h\wedge W\}. \nonumber
\end{align}
The induced $t$-slice metric is $\mu-F^{-1}(h+d\psi)\otimes(h+d\psi)$, which is positive precisely under \eqref{em:strict-sign}; compactness makes the positivity uniform on a tail.  Also $q^{tt}=-F+|h+d\psi|_\mu^2$ is uniformly negative. All inverses, lapse and shift, and the required derivatives are bounded. Although $u$ itself is unbounded, its positive-order derivatives are bounded. Agreement of full transverse jets makes the differences of $q,u,E,H,J$, and $A$ flat on the cylinder, meaning $O(e^{-Nt})$ with every fixed number of derivatives for every $N$.  Division by finitely many powers of $r$ preserves this statement. At fixed $t$, spatial differentiation in the original Gaussian-null coefficients is $\partial_a+r(\partial_a\psi)\partial_r$; thus the coordinate change retains all bounds. Since $\psi$ and all its derivatives are bounded on the compact section, $O(r^N)$ for every $N$ is still decay of every exponential order in $t$.

For clarity, the complete reduced equations used in the proof are
\begin{align}
dE&=-p\wedge E,&\quad \nabla^iE_i&=\langle H,J\rangle_q,\label{em:E}\\
dH&=-p\wedge H,& \nabla^iH_i&=-\langle E,J\rangle_q,\label{em:H}\\
dJ&=-2p\wedge J+4E\wedge H,& \nabla^iJ_i&=\langle p,J\rangle_q,\label{em:J}\\
\square_q u&=-|p|_q^2-\Lambda-|E|_q^2-|H|_q^2-\tfrac12|J|_q^2, \label{em:u}\\
\operatorname{Ric}(q)&=\nabla^2_q u+p\otimes p+2E\otimes E+2H\otimes H +\tfrac12J\otimes J\nonumber\\
&\hspace{13mm}+ \left(\Lambda-|E|_q^2-|H|_q^2-\tfrac12|J|_q^2\right)q. \label{em:q}
\end{align}
In \eqref{em:E}--\eqref{em:J}, all sources are algebraic in the fields and $p$; in particular no inverse Killing norm occurs.

Here is a derivation with the signs exposed.  Write $e^0=V\vartheta$, so that $de^0=p\wedge e^0+C$ and $\star_g\mathcal F=\star_q E+e^0\wedge H$.  Separating vertical and horizontal parts of $d\mathcal F=d\star_g\mathcal F=0$ gives
\begin{equation}
dE=-p\wedge E,\quad dB=-C\wedge E,\quad dH=-p\wedge H,\quad d\star_q E=-C\wedge H. \label{em:maxwell-derivation}
\end{equation}
Using $B=-\star_qH$, $C=-\star_qJ$ gives \eqref{em:E}--\eqref{em:H}. The connection-curvature identity is $dC=p\wedge C$.  The mixed Einstein component gives, with the indicated index order,
\begin{equation}
\nabla_i C^{ij}+2p_i C^{ij}=4E_i B^{ij}. \label{em:mixed}
\end{equation}
For example its normalization follows by varying $A$ in the reduced Einstein--Maxwell action: the connection term is $-\tfrac14\sqrt{|q|}V^3(dA)_{ij}(dA)^{ij}$, while variation of the horizontal Maxwell square contributes $4\sqrt{|q|}V B^{ij}(\iota_K\mathcal F)_j\delta A_i$. Equations $dC=p\wedge C$ and \eqref{em:mixed} are exactly \eqref{em:J}. Thus the vacuum divergence law has not been retained incorrectly in the charged problem.

For the remaining components the geometric identities are
\begin{align}
R_{00}(g)&=-\square_q u-|p|_q^2+\tfrac14 C_{ij}C^{ij},\label{em:ric00}\\
R_{ij}^{\rm hor}(g)&=\operatorname{Ric}_{ij}(q) -\nabla_i p_j-p_i p_j-\tfrac12C_{ik}C_j{}^k,\label{em:richor}\\
R_{0j}(g)&=-\tfrac12\bigl(\nabla^i C_{ij}+2p^i C_{ij}\bigr). \label{em:ricmixed}
\end{align}
These identities follow directly from the torsion-free connection forms
\begin{equation}
\omega^0{}_a=p_a e^0+\tfrac12 C_{ab}e^b,\qquad \omega^a{}_b=\bar\omega^a{}_b-\tfrac12C^a{}_b e^0, \qquad \omega^a{}_0=-p^a e^0-\tfrac12C^a{}_b e^b,
\end{equation}
by forming $d\omega+\omega\wedge\omega$ and contracting; $\bar\omega$ is the base connection.  Here the index $0$ denotes the unit spacelike fiber, not the timelike base coordinate.  In three Lorentzian dimensions $B_{ik}B_j{}^k=H_iH_j-q_{ij}|H|_q^2$ and $C_{ij}C^{ij}=-2|J|_q^2$.  Substitution in \eqref{em:equations} yields \eqref{em:u}--\eqref{em:q}.

Apply Lemma~\ref{pos:terminal-gauge} to $(q_1,u_1)$ and the general reference $(q_2,u_2)$, with $W=\operatorname{grad}_{q_2}u_2$. It supplies a flat perturbation $Y$ of the identity, constructed on a single late tail from terminal Cauchy problems.  Push the first solution forward by $Y$.  Suppressing the hats, its gauge is
\begin{equation}
\mathcal C(q_1,q_2)=\operatorname{grad}_{q_1}u_1-W, \quad \mathcal C(q_1,q_2)^k=q_1^{ij} (\Gamma(q_1)^k{}_{ij}-\Gamma(q_2)^k{}_{ij}). \label{em:gauge}
\end{equation}
Consequently $\operatorname{Ric}^{\mathcal C}(q_1) =\operatorname{Ric}(q_1)-\frac12\mathcal L_{\mathcal C}q_1$ has the right-hand side of \eqref{em:q} with $\nabla^2u_1$ removed and $\frac12\mathcal L_Wq_1$ added.  No Hessian of an unknown scalar remains. The matter terms in this reduced equation are algebraic.

Let $z=(q_1-q_2,u_1-u_2)$ and $U=(E_1-E_2,H_1-H_2,J_1-J_2)$.  In a fixed bundle connection $D$, subtraction of \eqref{em:u} and the reduced \eqref{em:q} has the form
\begin{equation}
q_1^{ij}D_iD_j z=\mathsf A^i D_i z+\mathsf Bz+\mathsf D U, \label{em:wave-difference}
\end{equation}
where the coefficient tensors and every fixed number of their derivatives are bounded on the tail.  Terms arising from a principal coefficient difference multiply known second derivatives of the second solution, hence belong to $\mathsf Bz$.  There is no derivative of $U$ in \eqref{em:wave-difference}.

For each one-form component $U^a$, $a\in\{E,H,J\}$, subtract \eqref{em:E}--\eqref{em:J} using $q_1$ as the divergence metric.  The result is
\begin{equation}
dU^a=\mathsf P_a(z,Dz,U),\qquad \operatorname{div}_{q_1}U^a=\mathsf Q_a(z,Dz,U), \label{em:curl-div-difference}
\end{equation}
with linear bounded-coefficient right-hand sides.  In particular, with $D=\nabla^{q_2}$, the exact change of divergence on any known reference one-form $X$ is
\[
(\operatorname{div}_{q_1}-\operatorname{div}_{q_2})X =(q_1^{ij}-q_2^{ij})D_iX_j -q_1^{ij}(\Gamma_1^k{}_{ij}-\Gamma_2^k{}_{ij})X_k.
\]
It has at most one derivative of $z$, and no derivative of $U$.  This elementary observation prevents the apparent derivative loss that would be introduced by applying another exterior derivative to the matter equations.

For a one-form $X$ with $dX=\mathsf P$ and $\operatorname{div}_{q_1}X=\mathsf Q$, use
\begin{equation}
T_{ij}[X]=X_iX_j-\tfrac12(q_1)_{ij}|X|_{q_1}^2, \qquad \nabla^iT_{ij}=\mathsf QX_j+X^i\mathsf P_{ij}. \label{em:oneform-energy}
\end{equation}
If $n$ is the future unit normal, then $T(n,n)=\frac12(|X(n)|^2+|X|_{S_t}^2)$, so it controls all components. Integrating this current over a compact-section slab and using bounded lapse, shift and deformation tensor of $n$ gives
\begin{equation}
|E_U'(t)|\leq C\{E_U(t)+E_z(t)\}. \label{em:matter-energy}
\end{equation}
The energies have the precise norm equivalences
\[
E_U\simeq\|U\|_{L^2}^2,\qquad E_z\simeq\|z\|_{H^1}^2+\|\partial_tz\|_{L^2}^2,
\]
with norms from a fixed positive product metric and bounded equivalence constants on the tail. Here $E_z$ is the positive Cauchy wave energy for \eqref{em:wave-difference}, including $\|z\|_{L^2}^2$.  The source estimate for that equation gives
\begin{equation}
|E_z'(t)|\leq C\{E_z(t)+E_U(t)\}. \label{em:metric-energy}
\end{equation}
Both bounds follow from Cauchy--Schwarz and $2ab\leq a^2+b^2$; no spatial boundary occurs.  More explicitly, the divergence theorem on any slab gives $|E_U(t_2)-E_U(t_1)|\leq C\int_{t_1}^{t_2}(E_U+E_z)\,dt$; the deformation term is bounded by $|T[U]|\,|\nabla n|$ and the source term by $|U|(|z|+|Dz|+|U|)$. The wave current has the same bound. Their constants depend only on finite coefficient bounds of the two known solutions on this tail and are independent of the terminal time.  In particular, for $E=E_z+E_U$,
\begin{equation}
E(t)\leq e^{C(T-t)}E(T). \label{em:combined-energy}
\end{equation}
All differences are flat, including after the wave-map change of variables. Thus $E(T)\leq C_Ne^{-2NT}$ for every $N$.  Choosing $2N>C$ and sending $T\to\infty$ yields $z=U=0$.

This comparison uses a metric gauge but no Maxwell potential or Lorenz condition.  The original and pulled-back solutions already satisfy the Maxwell and Einstein constraints.  No separate constraint-propagation assumption is introduced in this uniqueness argument.

In the original coordinates, equality of $q,u,J$ gives $dA_1=Y^*dA_2$.  The one-form $D=A_1-Y^*A_2$ is therefore closed and flat. Its global primitive is
\begin{equation}
f(t,x)=-\int_t^\infty D_t(s,x)\,ds. \label{em:primitive}
\end{equation}
Indeed $\partial_tf=D_t$ and $\partial_af=-\int_t^\infty\partial_sD_a\,ds=D_a$.  This proves exactness without assuming $H^1(S)=0$.  The map $\mathcal I(v,t,x)=(v+f(t,x),Y(t,x))$ identifies $g_1$ and $g_2$, and \eqref{em:regular-fields} identifies $\mathcal F_1$ and $\mathcal F_2$ as well. The factor $V$ and the connection may have exponential behavior, but their multiplication by flat differences preserves flatness.

Writing $x'=Y^S(t,x)$, the signed radial component is \[r'=r\exp[-(Y^0-t)+\psi(x')-\psi(x)].\]  All departures of $\mathcal I$ from the identity are smooth and flat in $r$ after extension by zero at $r=0$.  The two side maps consequently glue, with identity differential on the horizon, to a smooth local diffeomorphism on a uniform whole-section collar.  Gaussian-null transverse geodesics with the same initial points, normal vectors and affine parameters are unique.  The glued isometry is therefore the identity in the common marked normalization.  This proves Theorem~\ref{em:jet}.

\subsection{Charged Nariai rigidity}

For an application we take the spherical charged near-horizon geometry, which sits in the Reissner--Nordstr\"om--de Sitter family and has the charged Nariai limit \cite{Nariai1999,CardosoDiasLemos2004,Katona2024}. With $R>0$ its data are
\begin{equation}
\begin{aligned}
\mu_0&=R^2d\Omega_2^2,\quad h_0=0,\quad F_0=\frac{2\Lambda R^2-1}{R^2},\\
\Psi_0&=\frac{Q}{R^2},\quad Q^2=R^2(1-\Lambda R^2),\quad W_0=B_{S,0}=0.
\end{aligned}
\label{em:RN-data}
\end{equation}

\begin{lemma}[Smooth finite-order charged recursion]
\label{em:finite-order-recursion}
Fix the round data \eqref{em:RN-data} with $\tfrac12\leq x:=\Lambda R^2\leq1$. In constant-trace cut gauge every smooth solution has exactly the transverse jets of the one-parameter Reissner--Nordstr\"om--de Sitter model, including its $C=0$ near-horizon member. This is a statement about finite derivatives of each order and holds also at the endpoint $x=1/2$.
\end{lemma}
\begin{proof}
We detail the use of \cite[Propositions~1 and~2]{Katona2024}, so that analyticity is not hidden in the citation. The first-order trace is made constant by solving a scalar Poisson equation on $S^2$. At order $n\geq2$, the $rr$ equation determines the trace from the lower jets. The $ra$, $vr$, Maxwell and Bianchi equations determine the remaining weighted fields from the tracefree tensor $k$ and one-form $Z$ representing the difference from the model at this order. The two residual equations, obtained from the first-order system and its higher-order counterpart \cite[Eqs.~(21)--(22), (42), and (44)]{Katona2024}, are
\begin{align}
-\nabla^2 k&=-R^{-2}[2+\nu y]k-8\Psi_0\mathcal D Z,\nonumber\\
-\nabla^2 Z&=-R^{-2}[3+(\nu-2)y]Z+\Psi_0\operatorname{div}k, \label{em:residual-charged}
\end{align}
where $\nu=n(n+1)$, $y=2x-1$, and $\mathcal D Z=\nabla_{(a}Z_{b)}-\tfrac12\gamma_{ab}\operatorname{div}Z$. For $n=1$ the same equations hold after subtracting the pure-trace slope; the trace is the prescribed constant. The background charge satisfies $\Psi_0^2R^2=1-x$.

On a round two-sphere all one-forms are sums of electric and magnetic vector harmonics $V$ with $\lambda=\ell(\ell+1)$, $\ell\geq1$. All tracefree symmetric two-tensors are sums of $\mathcal D V$, $\ell\geq2$. There is no additional TT sector on $S^2$. The identities are
\[
-\nabla^2V=R^{-2}(\lambda-1)V,\quad -\nabla^2\mathcal D V=R^{-2}(\lambda-4)\mathcal D V,\quad \operatorname{div}\mathcal D V=\frac{2-\lambda}{2R^2}V.
\]
For $\ell\geq2$, writing $k=k_\ell\mathcal D V$ and $Z=z_\ell V$ reduces \eqref{em:residual-charged} to
\begin{equation}
\begin{pmatrix}
\lambda-2+\nu y&8\Psi_0R^2\\
\tfrac12\Psi_0(\lambda-2)&\lambda+2+(\nu-2)y
\end{pmatrix}
\binom{k_\ell}{z_\ell}=0. \label{em:charged-modal-matrix}
\end{equation}
Its determinant simplifies without dividing by $\Psi_0$ to
\begin{equation}
\lambda(\lambda-2)+2\nu\lambda y+\nu(\nu-2)y^2>0 \qquad(\lambda\geq6,\ \nu\geq2,\ y\geq0). \label{em:positive-determinant}
\end{equation}
For $\ell=1$ the tensor harmonic is zero, and the remaining coefficient is $4+(\nu-2)y>0$. Thus $k=Z=0$ at every order. The uncharged endpoint $x=1$ is included directly, with zero off-diagonal entries.

For each fixed $n$ this calculation only differentiates the exact equations finitely many times at the horizon. Taylor's theorem with smooth remainder, followed by the displayed elliptic identities, suffices; no infinite series is summed and no assumption about ambient electrostaticity is made. Katona's preliminary analytic and complete-generator setup is consequently not used for these finite-order identities. A local Killing flow and the whole-section Gaussian-null lemma supply their geometric setting. The induction determines the package $(\mu^{(n)},\beta^{(n+1)},\phi^{(n+2)},\Psi^{(n)},W^{(n)}, B_S^{(n)},Z^{(n-1)})$, hence every field-strength jet as well as every metric jet. This proves the claim.
\end{proof}

\begin{corollary}[Strict charged-Nariai regime]
\label{em:charged-nariai}
Let a smooth four-dimensional solution of \eqref{em:equations}, with $\Lambda>0$, contain a degenerate Killing horizon with an actual marked compact connected $S^2$ section and a nonvanishing normalized generator preserving the Maxwell field.  Suppose its round near-horizon data, after a fixed electromagnetic duality rotation if necessary, are exactly \eqref{em:RN-data}, with $\frac12<\Lambda R^2\leq1$.  Then on a two-sided neighborhood of the whole section the metric and Maxwell field are those of extremal Reissner--Nordstr\"om--de Sitter or its charged Nariai near-horizon solution. The isometry preserves the normalized generator and marking, with the original cut retained as a graph, and the fixed duality rotation is undone on both Maxwell fields.
\end{corollary}
\begin{proof}
Lemma~\ref{em:finite-order-recursion} determines every transverse coefficient after the constant-trace cut normalization. The exact model is Katona's Eq.~(51) \cite{Katona2024}, with slope parameter $C$. For completeness, put $x=\Lambda R^2$ and $b=RC/2$. Its full fields are
\begin{align}
g_C&=2\,dv\,d\rho+\rho^2F_C(\rho)dv^2 +(R+b\rho)^2d\Omega_2^2,\nonumber\\
F_C(\rho)&= \frac{2x-1+\frac23xC\rho+\frac1{12}xC^2\rho^2} {R^2(1+C\rho/2)^2},\qquad \mathcal F_C=\frac{Q}{(R+b\rho)^2}\,dv\wedge d\rho. \label{em:exact-spherical-model}
\end{align}
For $b\ne0$, these follow directly from the usual charged static metric with area radius $R+b\rho$, mass $R(1-2x/3)$ and charge $Q$, by rescaling its regular null time by $1/b$. The displayed formulas extend through $b=0$ and satisfy the equations there by continuity. It is smooth on a two-sided collar for every real $C$. Since $F_0>0$, Theorem~\ref{em:jet} identifies the smooth solution with this exact model. For $C=0$ it is charged Nariai; for $C\ne0$ the change $v'=2v/(RC)$, $\rho'=RC\rho/2$ and $r=R+\rho'$ gives extremal Reissner--Nordstr\"om--de Sitter. The comparison restores the originally marked cut as its corresponding graph. Neither analyticity nor ambient electrostaticity is used.
\end{proof}

The radius interval in this corollary is strict at $\Lambda R^2=1/2$. Katona's analytic theorem includes that ultracold endpoint and generic cold radii below it.  At the endpoint $F_0=0$ and the uniformly Lorentzian logarithmic cylinder used above is absent.  Below it $F_0<0$ and the orbit metric is Riemannian.  Neither case follows from this proof.  This distinction is independent of the formal resonance exceptions in the cold range.

\section{Rotating Nariai Horizons}\label{sec:rotating-nariai}
\label{sec:rotating}

Rotation supplies two distinct tests of the preceding results. First, we must find a spacelike foliation of the orbit cylinder even when the naive constant-$t$ slices fail. Second, to obtain a classification from intrinsic data alone, we must rule out every transverse tensor mode at every order, including nonaxisymmetric ones. The model is the known rotating Nariai limit of Kerr--de Sitter \cite{AnninosHartman2010}. We shall distinguish the graph statement, which holds on its full regular branch, from the stronger ambient classification, proved below on a smaller explicit interval. Earlier horizon and infinitesimal-deformation analyses \cite{KunduriLucietti2009,JezierskiKaminski2013,BahuaudEtAl2024} provide the natural comparison.

\subsection{Spacelike graphs of the rotating orbit space}

\begin{proposition}[Rotating Nariai graph slicing]\label{rotating:example}
Fix $\Lambda>0$. The cosmological double-root branch of four-dimensional Kerr--de Sitter is regular for
\begin{equation}
0\le x:=\Lambda a^2<21-12\sqrt3. \label{rotating:domain}
\end{equation}
Every corresponding rotating Nariai horizon has a smooth actual $S^2$ section, a nonzero normalized degenerate generator, and a global $\psi$ satisfying \eqref{graph:condition}. Consequently any two smooth vacuum germs with these marked horizon data and matching full transverse jets agree on a two-sided whole-section collar. For $a\ne0$ the generator is not hypersurface orthogonal.
\end{proposition}
\begin{proof}
Use the Kerr--de Sitter normalization of \cite[Eqs.~(2.2)--(2.4), (2.12)--(2.16)]{AnninosHartman2010}:
\[
\Delta_R=(R^2+a^2)(1-\Lambda R^2/3)-2MR.
\]
A positive double root $r_0$ obeys, with $s=r_0^2$,
\begin{align}
\Lambda s^2+(\Lambda a^2/3-1)s+a^2&=0,\nonumber\\
M&=r_0[1-\Lambda(2s+a^2)/3],\nonumber\\
D:=2\Lambda s+\Lambda a^2/3-1&=-\tfrac12\Delta_R''(r_0). \label{rotating:double}
\end{align}
The cosmological branch is the larger positive root and has $D>0$. Writing $y=\Lambda s$, its discriminant and root are
\[
\delta=(1-x/3)^2-4x,\qquad y=\tfrac12(1-x/3+\sqrt\delta),\qquad D=\sqrt\delta.
\]
Positivity of $y$ requires $x<3$; together with $\delta>0$ this gives exactly \eqref{rotating:domain}. The alternative discriminant interval $x>21+12\sqrt3$ has negative $y$ and gives no positive double root. For a check of the root ordering, put $z=a^2/s=x/y$. Then
\begin{equation}
y=\frac{3(1-z)}{3+z},\quad x=\frac{3z(1-z)}{3+z},\quad D=\frac{3-6z-z^2}{3+z},\quad 0\le z<z_*:=2\sqrt3-3. \label{rotating:z}
\end{equation}
In dimensionless radius $u=\sqrt\Lambda R$ the remaining quadratic factor of $\Delta_R$ is $u^2+2\sqrt y\,u-3x/y$. Its remaining nonnegative root (positive for $a\ne0$) is $-\sqrt y+\sqrt{y+3x/y}<\sqrt y$ precisely when $x<y^2$, equivalently $D>0$ on this branch. Thus $r_0$ is the outer/ cosmological double root and, for $a\ne0$, the other positive root is the inner one. The mass is positive because $M/r_0=(1+z)^2/(3+z)$. At $z=z_*$ the roots coalesce to a triple root and $D=0$; this endpoint is excluded.

Set
\begin{gather}
\Sigma=s+a^2\cos^2\theta,\qquad \Delta_\theta=1+\Lambda a^2\cos^2\theta/3,\qquad \Xi=1+\Lambda a^2/3,\nonumber\\
\Gamma=\Sigma/D,\quad A=\Sigma/\Delta_\theta,\quad B=\frac{(s+a^2)^2\Delta_\theta\sin^2\theta}{\Xi^2\Sigma},\quad k=-\frac{2ar_0\Xi}{D(s+a^2)}. \label{rotating:coefficients}
\end{gather}
The double-root scaling $R=r_0-\varepsilon X$, $\widehat t=(s+a^2)\tau/(D\varepsilon)$, $\widehat\phi=\Phi+[a\Xi/(s+a^2)]\widehat t$ in the exact Kerr--de Sitter metric gives
\begin{equation}
g=\Gamma(X^2d\tau^2-dX^2/X^2) +A\,d\theta^2+B(d\Phi+kX\,d\tau)^2. \label{rotating:nh}
\end{equation}
For instance the first squared term of Kerr--de Sitter supplies $\Sigma X^2d\tau^2/D$ and the radial term supplies $-\Sigma dX^2/(DX^2)$; the last squared term supplies the displayed $k$. This verifies the normalization directly from the parent metric.

On a punctured side let $\tau=v-1/X$, $\Phi=\varphi-k\log|X|$ and $r=\Gamma(\theta)X$. Direct substitution yields
\begin{gather}
g=2\,dv\,dr+r^2F\,dv^2+2r h\,dv+\gamma,\qquad \gamma=A\,d\theta^2+B\,d\varphi^2,\nonumber\\
h=-d\log\Gamma+\frac{kB}{\Gamma}d\varphi,\qquad F=\frac1\Gamma+\frac{k^2B}{\Gamma^2}. \label{rotating:gncdata}
\end{gather}
The coordinate $\varphi$ has period $2\pi$. At each pole, $B/A\sim\sin^2\theta$, $h_\varphi=O(\sin^2\theta)$, and $\Gamma=(s+a^2\cos^2\theta)/D$ is a positive smooth function of the height coordinate. Hence $\log\Gamma$ is a single-valued smooth function on the whole sphere. The Gaussian-null metric extends smoothly across $r=0$, where it remains Einstein by continuity; its determinant has Lorentzian sign. The section $v=r=0$ is spacelike and compact, $K=\partial_v$ is nonzero, and $g(K,\partial_r)=1$. The quadratic zero of $g(K,K)$ gives degeneracy.

Now take $\psi=\log\Gamma$. The exact part of $h$ cancels, and
\begin{equation}
F-|h+d\psi|_\gamma^2=\Gamma^{-1}=D/\Sigma>0. \label{rotating:graph-margin}
\end{equation}
Lemma~\ref{graph:lemma} and Theorem~\ref{strict:theorem} apply for every fixed parameter in \eqref{rotating:domain}. The uniform constants may deteriorate as $D\downarrow0$; there is no endpoint conclusion. Finally $h_\varphi$ vanishes at a pole but equals $-2ar_0(s+a^2)/(\Xi s^2)\ne0$ at the equator when $a\ne0$. Thus $dh$ is not identically zero, and
\[
\left.\iota_{\partial_r}(K^\flat\wedge dK^\flat)\right|_{TS^2}=r\,dh
\]
excludes hypersurface orthogonality on a whole-section collar.
\end{proof}

\begin{proposition}[Exact comparison with constant-$t$ slicing]
\label{rotating:old-domain}
Let $z=a^2/r_0^2$ and let $z_c$ be the unique root in $(0,z_*)$ of
\begin{equation}
R(z)=z^4-12z^3-74z^2+148z-47. \label{rotating:threshold-polynomial}
\end{equation}
The original condition $F>|h|_\gamma^2$ holds on the entire sphere exactly for $0\le z<z_c$. Numerically, $z_c=0.4046132094\ldots$, corresponding to $\Lambda a^2=0.2122720074\ldots$. The graph criterion instead holds for $0\le z<z_*=0.4641016151\ldots$, or $0\le\Lambda a^2<0.2153903092\ldots$.
\end{proposition}
\begin{proof}
Directly from \eqref{rotating:gncdata},
\begin{equation}
F-|h|^2=\frac D\Sigma- \frac{4a^4\Delta_\theta\sin^2\theta\cos^2\theta}{\Sigma^3}. \label{rotating:strict}
\end{equation}
Put $w=\cos^2\theta$. Its sign equals the sign of
\begin{align*}
P(w,z)={}&(3-6z-z^2)(1+zw)^2\\
&-4z^2[3+z+z(1-z)w]w(1-w).
\end{align*}
The endpoints $w=0,1$ are positive for $0\le z<z_*$. The minimum is positive at $z=0$ and negative for $z$ sufficiently close to $z_*$. Every first or later zero of the minimum in the open interval is an interior double root in $w$. Elimination gives the exact identity
\[
\operatorname{Res}_w(P,\partial_wP) =1728z^9(z-1)(z+1)^4R(z).
\]
On $[0,z_*]\subset[0,1/2]$, $R'(z)>65$, while $R(0)<0<R(z_*)$. It therefore has exactly one zero there. Continuity of the minimum, its endpoint signs and the resultant identity prove the claimed threshold without relying on a parameter scan. For an entirely rational strict separation, take $z=9/20$, so $y=11/23$, $x=99/460$, $D=13/460$. At $w=1/2$ one obtains
\[
(F-|h|^2)/\Lambda=-\frac{246346}{1294139}<0, \qquad (F-|h+d\psi|^2)/\Lambda=\frac{26}{539}>0.
\]
The numerical values above merely locate the algebraic endpoints.
\end{proof}

Thus complete transverse agreement implies smooth continuation on the entire regular branch. The remaining problem is to show when the horizon data themselves force that agreement. We solve it below on an explicit smaller interval.

\subsection{All-order transverse rigidity}\label{sec:small-rotation}
Keep the vacuum Kerr--de Sitter notation $(a,r_0,s,D,\Sigma, \Delta_\theta,\Xi,\Gamma,A,B,k)$ of the rotating example, with $s=r_0^2>0$ on the cosmological double-root branch.  Its horizon data are
\[
\gamma=A\,d\theta^2+B\,d\varphi^2,\qquad h=-d\log\Gamma+\frac{kB}{\Gamma}d\varphi,\qquad F=\Gamma^{-1}+\frac{k^2B}{\Gamma^2}.
\]
In this section a subscript $*$ denotes a chosen exact Kerr--de Sitter reference, not a new near-horizon solution.

\label{rot:general-transverse-section}

The following computation uses the full stationary Gaussian-null Ricci tensor, not a rotating ansatz for the perturbation. Let
\[
g_{\rm NH}=2\,dv\,dr+r^2F(x)\,dv^2+2r h_a(x)\,dv\,dx^a +\gamma_{ab}(x)\,dx^a dx^b
\]
be near-horizon data for $\Ric(g)=\Lambda g$. Here all contractions and covariant derivatives refer to $\gamma$, which need not be Einstein. Define
\begin{align}
(\Delta_L^\gamma k)_{ab} &=-\nabla^c\nabla_c k_{ab}+\Ric(\gamma)_a{}^c k_{cb} +\Ric(\gamma)_b{}^c k_{ac}-2R_{acbd}(\gamma)k^{cd}, \label{rot:general-Lichnerowicz}\\
(\operatorname{div}k)_a&=\nabla^b k_{ab},\qquad (dh)_{ab}=\nabla_a h_b-\nabla_b h_a.\nonumber
\end{align}
Parentheses on two indices include the factor $1/2$.

\begin{proposition}[General leading transverse operator]
\label{rot:general-indicial}
Suppose two smooth stationary Einstein metrics in common marked Gaussian-null coordinates have the same near-horizon data and agree through transverse order $n-1$, where $n\geq1$. Use monomial, rather than factorial, coefficients for their first difference:
\[
\delta\mu=r^n k+O(r^{n+1}),\qquad \delta\beta=r^{n+1}b+O(r^{n+2}),\qquad \delta\phi=r^{n+2}p+O(r^{n+3}).
\]
For $n\geq2$ the tensor $k$ is tracefree. For $n=1$ assume instead that the two first transverse traces are the same constant in the chosen cut gauge. Then
\begin{equation}
(n+1)b=n\,k(h,\cdot)-\operatorname{div}k, \label{rot:general-ra}
\end{equation}
and $k$ satisfies $\mathcal P_n k=0$, where
\begin{align}
(\mathcal P_n k)_{ab} ={}&(\Delta_L^\gamma k)_{ab}+(2n+1)h^c\nabla_c k_{ab} -2h^c\nabla_{(a}k_{b)c} +2h_{(a}(\operatorname{div}k)_{b)}\nonumber\\
&+2n k_{c(a}(dh)_{b)}{}^c +\{n(n+1)(F-|h|^2)+n\operatorname{div}h-2\Lambda\}k_{ab}. \label{rot:general-operator}
\end{align}
This is an elliptic operator on all tracefree symmetric tensors, with principal symbol $|\xi|_\gamma^2$ times the identity. It preserves the tracefree subbundle and depends smoothly on the near-horizon data. At $h=0$, $F=\Lambda$, $\Ric(\gamma)=\Lambda\gamma$, it is exactly $\Delta_L+[n(n+1)-2]\Lambda$.
\end{proposition}
\begin{proof}
We expand the raw Gaussian-null Ricci components \cite[Appendix~A, Eqs.~(57)--(60)]{KL}; no symmetry is imposed on $k$ or $b$. For clarity put $\tau=\tr_\gamma k$, $w_a=k_{ab}h^b$, and
\[
c^c{}_{ab}=\tfrac12\gamma^{cd} (\nabla_a k_{bd}+\nabla_b k_{ad}-\nabla_d k_{ab}),\qquad S_{ab}=\nabla_{(a}h_{b)}.
\]
The leading $rr$ residual is $-\frac12n(n-1)r^{n-2}\tau$. The leading $ra$ residual is
\begin{equation}
\frac n2r^{n-1} \{(n+1)b_a-nw_a+(\operatorname{div}k)_a -\nabla_a\tau+\tfrac12\tau h_a\}. \label{rot:raw-ra}
\end{equation}
Thus $\tau=0$ for $n\geq2$, and \eqref{rot:general-ra} follows. For $n=1$ we use the stated equality of the two normalized traces.

Here are the angular terms before the elimination of $b$, which also fix the sign and the factors in \eqref{rot:general-operator}. In the raw formula write
\[
Q_{ab}=2\nabla_{(a}\beta_{b)}+(\phi-|\beta|^2)\dot\mu_{ab}, \qquad J_a=\dot\beta_a-\beta^c\dot\mu_{ac}.
\]
Then
\begin{align*}
R_{ab}={}&\tfrac12\dot Q_{ab} +\tfrac14(\mu^{cd}\dot\mu_{cd})Q_{ab} +\tfrac12\nabla_c(\beta^c\dot\mu_{ab})+\Ric(\mu)_{ab}\\
&-\tfrac12J_aJ_b-\dot\mu_{c(a}\nabla^c\beta_{b)} +\tfrac12(|\beta|^2-\phi)\dot\mu_{ac}\dot\mu_{bd}\mu^{cd}.
\end{align*}
A dot means $\partial_r$. At the near-horizon background, $Q_{ab}=2rS_{ab}$ and $J_a=h_a$. Their variations, in the stated monomial convention, are
\[
\delta Q_{ab}=r^{n+1}\{2\nabla_{(a}b_{b)}-2c^c{}_{ab}h_c +n(F-|h|^2)k_{ab}\},\qquad \delta J_a=r^n\{(n+1)b_a-nw_a\}.
\]
Consequently the coefficient of $r^n$ in $\delta(\Ric(g)_{ab}-\Lambda\mu_{ab})$ is
\begin{align}
&(D\Ric)_\gamma(k)_{ab} +(n+1)(\nabla_{(a}b_{b)}-c^c{}_{ab}h_c) +\tfrac12n(n+1)(F-|h|^2)k_{ab}\nonumber\\
&\quad+\tfrac n2\tau S_{ab} +\tfrac n2\nabla_c(h^c k_{ab}) -h_{(a}\{(n+1)b_{b)}-nw_{b)}\} -n k_{c(a}\nabla^c h_{b)}-\Lambda k_{ab}. \label{rot:raw-angular}
\end{align}
For $\tau=0$ use
\[
(D\Ric)_\gamma(k)=\tfrac12\Delta_L^\gamma k +\nabla_{(a}(\operatorname{div}k)_{b)}, \qquad c^c{}_{ab}h_c=h^c\nabla_{(a}k_{b)c}-\tfrac12\nabla_h k_{ab}.
\]
Substitution of \eqref{rot:general-ra} cancels the entire second-derivative term $\nabla_{(a}(\operatorname{div}k)_{b)}$. The remaining coefficients are exactly one half of \eqref{rot:general-operator}.

For reference, the general-trace elimination replaces the right side of \eqref{rot:general-ra} by $n\,k(h,\cdot)-\operatorname{div}k+d\tau-\frac12\tau h$. Twice the angular residual then equals \eqref{rot:general-operator} plus
\[
\nabla^2\tau+(n-1)\tau S-3h_{(a}\nabla_{b)}\tau+\tau h\otimes h.
\]
This formula is not used to discard a trace equation: $\tau$ has already been fixed by $rr$ or by the first-order cut gauge.

The trace of $\Delta_L^\gamma k$ is $-\nabla^2\tau$. The two terms involving $h$ and $\operatorname{div}k$ have opposite traces, and the trace of the $dh$ term is zero by symmetry. Thus \eqref{rot:general-operator} preserves tracefree tensors. All added terms are at most first order, establishing ellipticity without a TT restriction.

The scalar coefficient is determined subsequently by $vr$. For $\tau=0$, its leading equation is
\begin{align}
0={}&(n+1)(n+2)(p-\langle h,b\rangle)+(n+1)k(h,h) +(n+1)\operatorname{div}b\nonumber\\
&\quad-\langle k,\nabla h\rangle -\langle h,\operatorname{div}k\rangle. \label{rot:general-vr}
\end{align}
The coefficient of $p$ is nonzero. In particular $k=0$ implies $b=p=0$.

Finally, positive-order background transverse coefficients multiply the new coefficient only at a strictly higher transverse weight in these components. Thus the calculation on $g_{\rm NH}$ is the leading difference calculation for arbitrary actual germs sharing their lower jets. The remaining Einstein components are satisfied by those actual germs; no existence from an arbitrary solution of this necessary indicial system is asserted. Factorial coefficients are recovered by multiplying $k,b,p$ by $n!,(n+1)!,(n+2)!$, respectively.
\end{proof}

For any fixed finite list of $n$, injectivity is stable under smooth perturbation of the horizon data near the round Nariai values. To see this, identify the varying tracefree bundles with a fixed one. At the round sphere the elliptic estimate and positivity give $\|k\|_{H^2}\leq C_n\|\mathcal P_n(0)k\|_{L^2}$. The operator difference tends to zero in $H^2\to L^2$ norm, so the right side absorbs that difference for sufficiently small perturbations. Taking the minimum of the resulting neighborhoods handles a finite list. The next estimate gives a stronger conclusion uniform in all positive integers $n$ for the rotating family considered here.

\begin{lemma}[Uniform coercivity of all transverse orders]
\label{rot:all-order-coercivity}
Let $m=\dim S$ and let $S$ be closed. On the tracefree tensor bundle define
\[
\mathcal A_\gamma k=\Ric(\gamma)_a{}^c k_{cb} +\Ric(\gamma)_b{}^c k_{ac}-2R_{acbd}k^{cd},\qquad \rho_\gamma=\min_{x\in S}\lambda_{\min}(\mathcal A_\gamma|_x).
\]
Set
\[
\begin{aligned}
c&=\min_S(F-|h|^2),\qquad H=\|h\|_{L^\infty},\\
\delta_h&=\|\operatorname{div}h\|_{L^\infty},\\
\eta&=\rho_\gamma+2c-2\Lambda-\tfrac12\delta_h-2(1+\sqrt m)^2H^2.
\end{aligned}
\]
If $c>0$ and $\eta>0$, then for every $n\geq1$ and every smooth tracefree $k$,
\begin{equation}
\langle\mathcal P_nk,k\rangle_{L^2} \geq\tfrac12\|\nabla k\|_{L^2}^2 +\{\eta+[n(n+1)-2]c\}\|k\|_{L^2}^2. \label{rot:all-order-energy}
\end{equation}
When $m=2$, the same estimate holds with the sharper constant
\[
\eta_2=\rho_\gamma+2c-2\Lambda-\tfrac12\delta_h-2H^2
\]
in place of $\eta$. In particular, either applicable positivity condition makes $\mathcal P_n$ injective at every positive order.
\end{lemma}
\begin{proof}
The estimate does not require $\mathcal P_n$ to be self-adjoint. Its apparently large first-order coefficients have exact cancellations:
\begin{align*}
\int_S\bigl((2n+1)\langle\nabla_hk,k\rangle +n(\operatorname{div}h)|k|^2\bigr) &=-\tfrac12\int_S(\operatorname{div}h)|k|^2,\\
k^{ab}k_{c(a}(dh)_{b)}{}^c&=0.
\end{align*}
The second equality contracts the symmetric tensor $k^2$ with the antisymmetric tensor $dh$. For the two remaining first-order terms, $|\operatorname{div}k|\leq\sqrt m\,|\nabla k|$ gives the pointwise bound
\[
\left|\left\langle-2h^c\nabla_{(a}k_{b)c} +2h_{(a}(\operatorname{div}k)_{b)},k\right\rangle\right| \leq2(1+\sqrt m)|h|\,|\nabla k|\,|k| \leq\tfrac12|\nabla k|^2+2(1+\sqrt m)^2H^2|k|^2.
\]
Integration by parts for $\Delta_L^\gamma=\nabla^*\nabla+\mathcal A_\gamma$ now gives \eqref{rot:all-order-energy}. All constants depend only on the fixed near-horizon data; none depend on $n$ or on unknown higher transverse jets.

For $m=2$, choose a normal orthonormal frame at a point and write $k_{11}=-k_{22}=u$, $k_{12}=v$, and $h=p\,e^1+q\,e^2$. With $u_j=\nabla_jk_{11}$ and $v_j=\nabla_jk_{12}$, the first-order tensor in the preceding estimate is exactly
\[
\mathcal B(k):=-2h^c\nabla_{(a}k_{b)c}+2h_{(a}(\operatorname{div}k)_{b)}
=2\begin{pmatrix}pv_2-qv_1&qu_1-pu_2\\ qu_1-pu_2&-pv_2+qv_1\end{pmatrix}.
\]
Thus $|\nabla k|^2=2\sum_{j=1}^2(u_j^2+v_j^2)$ and $|\mathcal B(k)|\leq2|h|\,|\nabla k|$. Young's inequality now gives
\[
|\langle\mathcal B(k),k\rangle|\leq\tfrac12|\nabla k|^2+2H^2|k|^2,
\]
which proves the stated two-dimensional improvement.
\end{proof}

\begin{proposition}[An explicit rotating interval]
\label{rot:explicit-coercivity}
For the vacuum rotating Nariai data, let $s=r_0^2$ and $z=a^2/s$ on the regular cosmological double-root branch. If
\begin{equation}
0\leq z\leq\frac18, \quad\text{equivalently}\quad 0\leq\Lambda a^2\leq\frac{21}{200}, \label{rot:explicit-interval}
\end{equation}
then $c>2\Lambda/3$ and $\eta_2>5\Lambda/24$ in Lemma~\ref{rot:all-order-coercivity}. Hence every $\mathcal P_n$, $n\geq1$, is injective on the full tracefree tensor space, with no axisymmetry assumption on a perturbation.
\end{proposition}
\begin{proof}
Put
\[
d=\frac{\Lambda a^2}{3}=\frac{z(1-z)}{3+z},\qquad s\Lambda=\frac{3(1-z)}{3+z},\qquad D=\frac{3-6z-z^2}{3+z},\qquad w=\cos^2\theta.
\]
Direct substitution in the explicit Gaussian-null data gives
\begin{align}
|h|^2&=\frac{4z(1+dw)(1-w)}{s(1+zw)^2},\nonumber\\
\operatorname{div}h &=-\frac{2z}{s}\frac{d}{dt} \left[\frac{(1+dt^2)(1-t^2)t}{(1+zt^2)^2}\right], \qquad t=\cos\theta.\label{rot:explicit-norms}
\end{align}
The interval in \eqref{rot:explicit-interval} gives $0\leq d\leq1/24$, $s\Lambda\geq21/25$, and $D>0$. It follows that
\[
\frac{H^2}{\Lambda} \leq\frac{4(1/8)(1+1/24)}{21/25}=\frac{625}{1008}<\frac58.
\]
For the divergence bound, set $A(w)=(1+dw)/(1+zw)^2$. Since $d\leq z/3$ and $zw\leq1$, $0\leq A\leq1+d$ and $|A'|\leq2z$. Therefore
\[
\left|\frac{d}{dt}\{t(1-t^2)A(t^2)\}\right| \leq2(1+d)+z\leq(1+d)(2+3z).
\]
Using \eqref{rot:explicit-norms} yields the rational bound
\[
\frac{\delta_h}{\Lambda} \leq\frac{2(1/8)(1+1/24)(2+3/8)}{21/25}=\frac{11875}{16128}<\frac34.
\]
Furthermore
\[
|d\log\Gamma|^2 =\frac{zw}{1+zw}|h|^2\leq zH^2\leq\frac{625}{8064}\Lambda.
\]
The function $(3-6z-z^2)/[3(1-z^2)]$ is decreasing on $0\leq z\leq1/8$, so
\[
\frac{D}{s\Lambda(1+z)}=\frac{3-6z-z^2}{3(1-z^2)}\geq\frac{143}{189}.
\]
Since $F-|h|^2=D/[s(1+zw)]-|d\log\Gamma|^2$, we obtain
\[
c\geq\left(\frac{143}{189}-\frac{625}{8064}\right)\Lambda=\frac{2347}{3456}\Lambda>\frac23\Lambda.
\]
In two dimensions $\mathcal A_\gamma=4K$ on tracefree tensors. The trace of the vacuum horizon equation reads $2K=2\Lambda+\frac12|h|^2-\operatorname{div}h$, so $\rho_\gamma\geq4\Lambda-2\delta_h$. The improved constant in Lemma~\ref{rot:all-order-coercivity} consequently satisfies
\[
\eta_2\geq2\Lambda+2c-\tfrac52\delta_h-2H^2
>\left(2+\frac43-\frac{15}{8}-\frac54\right)\Lambda=\frac5{24}\Lambda.
\]
The conversion to the second interval in \eqref{rot:explicit-interval} uses $\Lambda a^2=3z(1-z)/(3+z)$, which is strictly increasing on the regular cosmological branch; its derivative is $3D/(3+z)>0$.
\end{proof}

\begin{lemma}[Constant-trace cuts]\label{rotating:cut-gauge}
Let $k^{(1)}$ be the first transverse section tensor of a smooth degenerate Killing horizon with the preceding data, and let $\tau=\tr_\gamma k^{(1)}$.  There is a smooth change of cut for which this trace is the constant
\begin{equation}
\bar\tau=\frac{\int_{S^2}\Gamma\tau\,d\mu_\gamma} {\int_{S^2}\Gamma\,d\mu_\gamma}. \label{rotating:weighted-trace}
\end{equation}
The change is unique modulo a constant translation along the generator. The horizon data $(\gamma,h,F)$ and the number $\bar\tau$ do not change. No symmetry assumption on $k^{(1)}$ or on the new cut is needed.
\end{lemma}
\begin{proof}
In a Gaussian-null collar let $K=\partial_v$, $N=\partial_r$ at the original cut, and let $e_a$ be the section coordinate vectors.  On the cut $v=f(x)$ its tangents and normalized transverse null normal are
\[
e_a^f=e_a+f_aK,\qquad N_f=N-\nabla^\gamma f-\tfrac12|df|_\gamma^2K.
\]
At the horizon, Killing invariance and degeneracy give
\[
\nabla_{e_a}K=-\tfrac12h_aK,\quad \nabla_KN=\tfrac12h^ae_a,\quad \nabla_{e_a}N=\tfrac12h_aN+\tfrac12(k^{(1)})_a{}^be_b, \quad\nabla_KK=0.
\]
Evaluating $2g(\nabla_{e_a^f}N_f,e_b^f)$ therefore gives the exact finite-cut formula
\begin{equation}
k^{(1)}_f=k^{(1)}-2\nabla^2f+2h_{(a}\nabla_{b)}f, \qquad \tau_f=\tau-2Lf,\quad L=\Delta_\gamma-h^a\nabla_a. \label{rotating:finite-cut}
\end{equation}
Every quadratic $df$ term has zero tangential contraction; the displayed formula is not merely an infinitesimal gauge transformation.  The induced metric is unchanged because $K$ is null and orthogonal to the horizon. The rotation one-form is unchanged because $\nabla_{e_a+f_aK}K=\nabla_{e_a}K$.  Thus $h$ is unchanged, and the vacuum horizon equation $F=\Lambda+\tfrac12|h|^2-\tfrac12\operatorname{div}h$ also preserves $F$.

For the rotating Nariai data the operator has the useful form
\begin{equation}
L=\Gamma^{-1}\operatorname{div}_\gamma(\Gamma\nabla) -\frac{k}{\Gamma}\partial_\varphi. \label{rotating:weighted-drift}
\end{equation}
The smooth vector $\partial_\varphi$ is Killing and preserves $d\mu_\gamma$.  Hence $L^*\Gamma=0$ and, for real $f$,
\[
\int\Gamma f Lf\,d\mu_\gamma =-\int\Gamma|df|^2\,d\mu_\gamma.
\]
On the space $\int\Gamma f=0$, the bilinear form of $-\Gamma L$ is coercive by the weighted Poincar\'e inequality; its axial first-order term is skew.  Lax--Milgram and elliptic regularity give a unique smooth solution of
\[
Lf=\tfrac12(\tau-\bar\tau),\qquad \int\Gamma f=0.
\]
The same identity shows that the kernel of $L$ consists of constants and that \eqref{rotating:weighted-trace} is cut invariant.  A bounded cut can be constructed in the auxiliary stationary product collar if the original Killing flow is not complete.  The original marked cut is then kept as its graph in the new coordinates.
\end{proof}

\begin{lemma}[Reference trace on the regular branch]
\label{rotating:reference-slope}
There is a smooth exact Kerr--de Sitter reference with the displayed horizon data whose constant-trace normalized first derivative has trace
\begin{equation}
\tau_*(a)=\frac{4r_0}{s+a^2/3}>0. \label{rotating:reference-trace}
\end{equation}
This holds throughout the regular cosmological double-root branch. At $a=0$ the normalization can be chosen so that $k_*^{(1)}=2\gamma/r_0$.  Reversing the reference generator and transverse affine orientation reverses the sign of both $k_*^{(1)}$ and $\tau_*$; only the fixed choice in \eqref{rotating:reference-trace} is used below.
\end{lemma}
\begin{proof}
Use outgoing regular Kerr--de Sitter coordinates $(w,R,\theta,\phi)$:
\begin{align}
g_*={}&-\frac{\Delta_R}{\Sigma_R} \left(dw-\frac{a\sin^2\theta}{\Xi}d\phi\right)^2 -2\left(dw-\frac{a\sin^2\theta}{\Xi}d\phi\right)dR +\frac{\Sigma_R}{\Delta_\theta}d\theta^2\nonumber\\
&+\frac{\Delta_\theta\sin^2\theta}{\Sigma_R} \left(a\,dw-\frac{R^2+a^2}{\Xi}d\phi\right)^2, \qquad \Sigma_R=R^2+a^2\cos^2\theta. \label{rotating:regular-reference}
\end{align}
It is smooth through the double root.  Put $\phi=\varphi+\Omega_Hw$, $\Omega_H=a\Xi/(s+a^2)$, and choose the reference generator $K=-\partial_w$ in these corotating coordinates.  On the actual cut $w=0,R=r_0$ the induced metric is precisely $\gamma$.  Write $c=\Sigma/(s+a^2)$ and $d=a\sin^2\theta/\Xi$.  The nonzero mixed horizon components are $g_{wR}=-c$ and $g_{R\varphi}=d$.  Solving $g(N,K)=1$, $g(N,TS)=0$ and $g(N,N)=0$ gives
\begin{equation}
N^R=\frac{s+a^2}{\Sigma},\qquad N^\varphi=-\frac{a\Xi}{(s+a^2)\Delta_\theta},\qquad N^w=-\frac{a^2\sin^2\theta}{2\Sigma\Delta_\theta},\qquad N^\theta=0. \label{rotating:reference-normal}
\end{equation}
All these define smooth fields at the poles.  The horizon rotation form can be checked directly: $h_\theta=-\partial_\theta\log\Sigma$ and
\[
h_\varphi=-\frac{2ar_0(s+a^2)\Delta_\theta\sin^2\theta} {\Xi\Sigma^2}=\frac{kB}{\Gamma}.
\]
For example the second identity follows from $h_\varphi=-c^{-1}\partial_Rg_{w\varphi}|_{r_0}$; the first follows from $g_{wR}=-c$.  The constant reversal of $K$ does not change this intrinsic rotation form.  The horizon equation then also gives the displayed $F$. Thus the reference has exactly the required intrinsic data.

Only the radial part of $N$ contributes to the trace of $\mathcal L_Ng|_{TS}$: the angular part is axial, its derivatives have no diagonal contraction, and $g(K,TS)=0$ on the horizon.  For a surface of constant $w,R$, direct multiplication gives
\[
\det\gamma_R=\frac{\sin^2\theta}{\Xi^2} \left[(R^2+a^2)^2- \frac{\Delta_Ra^2\sin^2\theta}{\Delta_\theta}\right].
\]
Since $\Delta_R(r_0)=\Delta_R'(r_0)=0$,
\[
\partial_R\log\det\gamma_R|_{r_0}=\frac{4r_0}{s+a^2}, \qquad \tau_{\rm raw}=N^R\partial_R\log\det\gamma_R|_{r_0} =\frac{4r_0}{\Sigma}.
\]
The area density is $(s+a^2)\sin\theta/\Xi$.  Its weighted average with $\Gamma=\Sigma/D$ is \eqref{rotating:reference-trace}, since the average of $\cos^2\theta$ for this density is $1/3$.

The reference normalization is also explicit.  A cut function is
\begin{equation}
f_*(\theta)=\frac{r_0}{\Lambda(s+a^2/3)}\log\Delta_\theta, \label{rotating:reference-cut}
\end{equation}
up to an additive constant.  In $x=\cos\theta$, for an axisymmetric function,
\[
Lf=\Sigma^{-1}\partial_x \left[(1-x^2)\Delta_\theta\partial_xf\right].
\]
Substitution gives $2Lf_* =\tau_{\rm raw}-\tau_*$.  This cut is smooth at both poles and as $a\to0$.  At $a=0$, set $v=-w$ and $r=R-r_0$; then $g_{vr}=1$, $\mu=(1+r/r_0)^2\gamma$, and $k_*^{(1)}=2\gamma/r_0$.  The Gaussian-null construction from \eqref{rotating:reference-normal}, followed by this cut change, supplies a smooth exact reference for every parameter in the regular branch.
\end{proof}

Write this normalized reference as
\[
g_*=2dv\,dr+r^2F_*(r,x)dv^2+2rh_*(r,x)dv\,dx+ \mu_*(r,x).
\]
For $\lambda\ne0$, pull it back by $(v,r,x)\mapsto(v/\lambda,\lambda r,x)$.  The resulting exact family has coefficient functions
\begin{equation}
(F_\lambda,h_\lambda,\mu_\lambda)(r,x) =(F_*,h_*,\mu_*)(\lambda r,x), \qquad k_\lambda^{(1)}=\lambda k_*^{(1)}. \label{rotating:scaling-family}
\end{equation}
It extends smoothly at $\lambda=0$ to the exact rotating Nariai metric. Every member has the same horizon data, and its normalized first trace is $\lambda\tau_*$.  Negative $\lambda$ is allowed: the reference generator and transverse orientation are then both reversed by the displayed isometry.  The comparison generator in the pulled-back coordinates remains $\partial_v$.

\begin{theorem}[Smooth ambient rigidity in an explicit rotation interval]
\label{rotating:ambient-classification}
Fix $\Lambda>0$ and the vacuum rotating Nariai horizon data on the regular cosmological Kerr--de Sitter double-root branch.  Suppose
\begin{equation}
0\leq z:=\frac{a^2}{r_0^2}\leq\frac18. \label{rotating:ambient-range}
\end{equation}
Equivalently on this branch, $0\leq\Lambda a^2\leq21/200$: the double-root identity gives $\Lambda a^2=3z(1-z)/(3+z)$, whose derivative is positive on the regular cosmological branch $3-6z-z^2>0$. Let a smooth four-dimensional solution of $\Ric(g)=\Lambda g$ have an actual compact marked $S^2$ section of a degenerate Killing horizon, a nonzero normalized generator, and these intrinsic horizon data.  After the constant-trace cut normalization of Lemma~\ref{rotating:cut-gauge}, the germ is smoothly isometric on a two-sided neighborhood of the whole section to precisely the member $g_\lambda$ in \eqref{rotating:scaling-family} with
\begin{equation}
\lambda=\frac{\bar\tau}{\tau_*(a)}. \label{rotating:ambient-parameter}
\end{equation}
The isometry preserves the normalized generator and the normalized marking.  The original cut is retained as a graph in this model. If $\lambda=0$, the ambient germ is rotating Nariai; if $\lambda\ne0$, it is locally Kerr--de Sitter.  No axial symmetry, analyticity, staticity, or prescribed higher transverse jet is assumed for the ambient germ.
\end{theorem}
\begin{proof}
Normalize the cut by Lemma~\ref{rotating:cut-gauge} and choose $\lambda$ by \eqref{rotating:ambient-parameter}, which is possible and unique by \eqref{rotating:reference-trace}.  The first section-tensor difference between the given germ and $g_\lambda$ has zero trace.  The first-order $ra$ and $vr$ equations express the differences of the other first transverse coefficients in terms of this tensor.  Their angular residual is $\mathcal P_1k=0$ from Proposition~\ref{rot:general-indicial}. The estimate of Proposition~\ref{rot:explicit-coercivity} is strictly positive on every trace-free symmetric tensor on $S^2$, so $k=0$. The algebraic elimination then identifies the whole first transverse package.

Inductively suppose the packages through order $n-1$ agree, with $n\geq2$. The leading $rr$ equation makes the order-$n$ section-tensor difference trace-free; the $ra$ and $vr$ equations again determine the other new coefficients.  Proposition~\ref{rot:general-indicial} gives $\mathcal P_nk=0$. Lemma~\ref{rot:all-order-coercivity} and Proposition~\ref{rot:explicit-coercivity} exclude this kernel for every $n\geq1$ under \eqref{rotating:ambient-range}.  Thus the full packages agree at order $n$.  This is an induction for two existing Einstein germs, so no Fredholm range condition or formal-to-exact existence assertion is being inserted.

All transverse jets now agree.  The orbit cylinder has a uniform spacelike foliation: throughout the regular branch $F-|h+d\log\Gamma|^2=\Gamma^{-1}>0$; in the smaller interval \eqref{rotating:ambient-range} even $F-|h|^2>0$. The smooth continuation theorem therefore gives the asserted isometry on both sides of the whole normalized section.  The cut construction and the reference family are exact, so this is an ambient classification, not a classification of formal series.  Pulling the original marked cut back as its graph restores the original whole-section germ.
\end{proof}

The two parameter ranges in this argument answer different questions. The graph construction works throughout the regular branch; the estimate for $\mathcal P_n$ supplies ambient classification for $z\leq1/8$. Intrinsic horizon symmetry results \cite{DunajskiLucietti,Colling2025}, higher-dimensional axial rigidity \cite{HollandsIshibashi2009}, and subextremal Kerr--de Sitter uniqueness \cite{Fang2025} concern different data or hypotheses. Here no symmetry is imposed on the transverse tensors. The parameter $\lambda$ records the chosen horizon marking: nonzero values describe locally Kerr--de Sitter metrics, not a new family of unmarked solutions.

\subsection{Charged rotating Nariai data}\label{em:charged-rotating}

The graph-slicing criterion also applies to the regular charged rotating Nariai branch.  This is an application to an existing exact family \cite[Appendix~B]{AnninosHartman2010}; the electromagnetic normalization is that of \eqref{em:equations}.  Let $a$ and $Q$ be real, $s=r_0^2>0$, and set
\begin{align}
Q^2&=s-a^2-\Lambda s^2-\tfrac\Lambda3a^2s,\nonumber\\
M&=r_0[1-\tfrac\Lambda3(2s+a^2)],\qquad D=2\Lambda s+\tfrac\Lambda3a^2-1>0. \label{em:charged-double-root}
\end{align}
They are exactly the double-root equations for $\Delta_R=(R^2+a^2)(1-\Lambda R^2/3)-2MR+Q^2$ on the cosmological branch.  In terms of $x=\Lambda a^2$ and $e=\Lambda Q^2$, its component connected to the uncharged Nariai solution is
\begin{equation}
0\leq x<21-12\sqrt3,\qquad 0\leq e<\tfrac14(1-x/3)^2-x, \label{em:charged-domain}
\end{equation}
with $\Lambda s=\frac12[1-x/3+\sqrt{(1-x/3)^2-4(x+e)}]$ and $D=\sqrt{(1-x/3)^2-4(x+e)}$. Indeed $\Delta_R(r_0)=\Delta'_R(r_0)=0$ gives \eqref{em:charged-double-root}; the upper root has $D>0$. On \eqref{em:charged-domain}, $s>0$ and $M>0$: writing $S=\Lambda s$, we have $0<S\leq1-x/3$ and hence $2S+x<3$. The remaining roots follow from
\[
\Delta_R=-\frac\Lambda3(R-r_0)^2 \left(R^2+2r_0R-\frac{3(a^2+Q^2)}{\Lambda s}\right).
\]
The remaining nonnegative root (positive when $a^2+Q^2>0$) is smaller than $r_0$ precisely when $a^2+Q^2<\Lambda s^2$, which is equivalent to $D>0$ by \eqref{em:charged-double-root}. Thus the double root is the outer/cosmological one throughout this component. The strict upper boundary has $D=0$, where this normalization of the Nariai limit degenerates.  No conclusion below includes it.

Use $\Sigma,\Delta_\theta,\Xi,\Gamma,A,B,k$ from the rotating vacuum construction, with this $s,D$. The same Gaussian-null data obey
\begin{equation}
\gamma=A\,d\theta^2+B\,d\varphi^2,\quad h=-d\log\Gamma+\frac{kB}{\Gamma}d\varphi,\quad F=\frac1\Gamma+\frac{k^2B}{\Gamma^2}. \label{em:charged-gncdata}
\end{equation}
An explicitly regular electric Maxwell field is
\begin{align}
\mathcal F&=\Psi\,dv\wedge d\rho+\rho\,dv\wedge d\Psi +df\wedge d\varphi,\nonumber\\
\Psi&=\frac{Q(s-a^2\cos^2\theta)}{\Sigma^2},\qquad f=\frac{Qa r_0\sin^2\theta}{\Xi\Sigma}. \label{em:charged-NH-field}
\end{align}
Here $f\,d\varphi$ extends smoothly across both poles and $\Psi$ is a smooth function of $\cos^2\theta$. A magnetic charge is obtained by a fixed electromagnetic duality rotation, replacing $Q^2$ in \eqref{em:charged-double-root} by the sum of squared electric and magnetic charges. The field strength is global even when a magnetic potential is not.

To check \eqref{em:charged-NH-field}, start from the usual electric Kerr--Newman--de Sitter potential
\[
-\frac{QR}{\Sigma_R}\left(d\widehat t-\frac{a\sin^2\theta}{\Xi}\,d\widehat\varphi\right)
\]
and subtract its constant horizon value. Apply the double-root scaling and corotating angle
\[
R=r_0-\epsilon x,\qquad \widehat t=\frac{s+a^2}{D\epsilon}T,\qquad \widehat\varphi=\Phi+\frac{a\Xi}{s+a^2}\widehat t.
\]
The limit has potential $f\,d\Phi+e(\theta)x\,dT$ with $e=-Q(s-a^2\cos^2\theta)/(D\Sigma)$. Under $T=v-1/x$, $\Phi=\varphi-k\log|x|$, $\rho=\Gamma x$, the remaining singular term is $(e-kf)d\log|x|=-Q(s-a^2)[D(s+a^2)]^{-1}d\log|x|$. Its coefficient is constant, so its exterior derivative vanishes. The regular potential $f\,d\varphi-\rho\Psi\,dv$ gives \eqref{em:charged-NH-field} on both sides.

Finally $\psi=\log\Gamma$ is a global smooth function on the actual sphere and
\begin{equation}
F-|h+d\psi|_\gamma^2=\Gamma^{-1}>0. \label{em:charged-graph}
\end{equation}
Thus Theorem~\ref{em:jet}, with the graph-slicing hypothesis, applies at every fixed parameter in \eqref{em:charged-domain}. Constants need not be uniform as $D\downarrow0$. This assertion compares solutions with matching complete transverse jets; it does not classify arbitrary ambient solutions with these rotating near-horizon data.

\section{Discussion}\label{Sec:Discussion}

The transverse field equations and smooth continuation answer distinct questions. The former constrain the horizon data order by order; the latter must exclude a smooth difference invisible to every transverse derivative. A spacelike slicing of the Killing-orbit cylinder makes that exclusion possible without analyticity. Once continuation is available, the classification problem lies in the transverse equations on the compact section. In the unwarped case this leads to an explicit finite matching order, while the finite-regularity results require a separate threshold obtained from the decay estimates.

The spectral picture distinguishes genuine moduli from possible formal ones. Parallel first-order tensors give actual families of Einstein germs: the multiwarped construction realizes each of them by a convergent solution. By contrast, a nonparallel first-order kernel or a higher-order resonance only signals a possible departure from the model. Its nonlinear compatibility and realization remain to be established \cite{MarsI,MarsII}. The K\"ahler--Einstein examples show, in particular, that negative sectional curvature is no obstacle to rigidity when the full tensor spectrum has the required gap.

Rotation brings the independence of the two conditions into focus. For rotating Nariai horizons the orbit graph exists throughout the regular cosmological double-root branch. The two-dimensional structure of tracefree tensors sharpens the all-order coercivity estimate and gives ambient classification for $a^2/r_0^2\leq1/8$, without imposing symmetry on the ambient spacetime. This is a sufficient interval, not a known threshold for loss of rigidity. Determining the kernels at larger rotation remains an immediate problem. The analogous all-order classification for charged rotating horizons is also open, even though their regular orbit geometry admits a spacelike graph.

The graph criterion itself may be restrictive. It is equivalent to stable causality of the limiting stationary orbit cylinder; when it fails, a different continuation mechanism would be needed. This is especially relevant for horizons with nonpositive cosmological constant, whose transverse spectral problems are already markedly different \cite{KL,Wylie2023}. The sharp finite weighted matching order and the least metric regularity needed for continuation are likewise open analytic questions.

Our conclusions concern germs near a marked compact horizon section. Extending such a germ into an exterior region introduces asymptotic conditions and, for de Sitter black holes, other horizons \cite{BorghiniChruscielMazzieri2023,Fang2025}. The local rigidity results separate that global problem from the question of whether the near-horizon expansion determines the smooth spacetime in the first place.

\section*{Statements and Declarations}

\paragraph{Funding}
No funds, grants, or other support were received.

\paragraph{Competing interests}
The author has no relevant financial or non-financial interests to disclose.

\paragraph{Author contributions}
Ye Zhou: Conceptualization, Methodology, Formal analysis, Investigation, Writing -- original draft, Writing -- review and editing.

\paragraph{Data availability}
No datasets were generated or analyzed during this study.

\paragraph{Use of generative AI}
OpenAI ChatGPT was used to assist with literature organization, manuscript preparation, and the exploration and checking of mathematical arguments. The author reviewed the AI-assisted material and takes full responsibility for the manuscript.

\bibliography{reference}

\end{document}